\RequirePackage{scrlfile}
\PreventPackageFromLoading[]{natbib}

\documentclass[nonatbib,5p]{elsarticle}
\makeatletter

\let\c@author\relax
\makeatother

\usepackage{dblfloatfix}    
\usepackage{etoolbox,siunitx}
\robustify\bfseries

\usepackage{booktabs}
\usepackage{pdfpages} 

\usepackage{url}
\usepackage{caption}
\usepackage{subcaption}
\usepackage[inline]{enumitem}
\newlist{mylist}{enumerate*}{1}
\setlist[mylist]{label=(\roman*)}
\usepackage{color}
\usepackage{graphicx}
\usepackage{algorithm}
\usepackage[noend]{algpseudocode}
\usepackage{multirow}
\usepackage{textcomp} 
\usepackage[none]{hyphenat} 

\usepackage{amsfonts}       
\usepackage{amsmath, bm}    
\usepackage[utf8]{inputenc}
\usepackage[english]{babel}

\RequirePackage[					
	backend=biber,		
	bibencoding=utf8,				
	style=authoryear,		
	natbib=false,					
	dashed=false,                   
	hyperref=true,					
	backref=false,					
	isbn=false,						
	url=false,						
	doi=true,						
	urldate=long,					
	uniquename=false,               
	uniquelist=minyear,
	maxbibnames=5,%
	minbibnames=1,%
	maxcitenames=2,%
	mincitenames=1%
]{biblatex}
\definecolor{mygreen}{rgb}{0.06, 0.48, 0.023}

\usepackage{hyperref} 
\hypersetup{					
	plainpages=false,			
	colorlinks=true,			
	pdfborder={0 0 0},			
	breaklinks=true,			
	bookmarksnumbered=true,		%
	bookmarksopen=true			%
}
    
\hypersetup{
    linkcolor = black,
	citecolor = mygreen,
	urlcolor = blue,
}

\usepackage{fancyhdr}
\begin{document}
\sloppy 

\hypersetup{allcolors = mygreen}

\title{Mapping horizontal and vertical urban densification in Denmark with Landsat time-series from 1985 to 2018: a semantic segmentation solution}

\journal{Remote Sensing of Environment (accepted)}
\author[add1,add2,add3]{Tzu-Hsin Karen Chen\corref{corr}}
\cortext[corr]{Corresponding author}
\ead{thc@envs.au.dk}

\author[add4]{Chunping Qiu}
\author[add4,add5]{Michael Schmitt}
\author[add4,add6]{Xiao Xiang Zhu}
\author[add1,add2]{Clive E. Sabel}
\author[add3]{Alexander V. Prishchepov}

\address[add1]{Department of Environmental Science, Aarhus University, Frederiksborgvej 399, DK-4000 Roskilde, Denmark}
\address[add2]{Danish Big Data Centre for Environment and Health (BERTHA), Aarhus University, DK-4000 Roskilde, Denmark}
\address[add3]{Department of Geosciences and Natural Resource Management (IGN), University of Copenhagen, DK-1350 København K, Denmark}
\address[add4]{Signal Processing in Earth Observation (SiPEO), Technical University of Munich (TUM), 80333 Munich, Germany}
\address[add5]{Department of Geoinformatics, Munich University of Applied Sciences, 80333 Munich, Germany}
\address[add6]{Remote Sensing Technology Institute (IMF), German Aerospace Center (DLR), 82234 Wessling}

\begin{abstract}

Landsat imagery is an unparalleled freely available data source that allows reconstructing land-cover and land-use change, including horizontal and vertical urban form. This paper addresses the challenge of using Landsat data, particularly its 30m spatial resolution, for monitoring three-dimensional urban densification. Unlike conventional convolutional neural networks (CNNs) for scene recognition resulting in resolution loss, the proposed semantic segmentation framework provides a pixel-wise classification and improves the accuracy of urban form mapping. We compare temporal and spatial transferability of an adapted DeepLab model with a simple fully convolutional network (FCN) and a texture-based random forest (RF) model to map urban density in the two morphological dimensions: horizontal (compact, open, sparse) and vertical (high rise, low rise). We test whether a model trained on the 2014 data can be applied to 2006 and 1995 for Denmark, and examine whether we could use the model trained on the Danish data to accurately map ten other European cities. Our results show that an implementation of deep networks and the inclusion of multi-scale contextual information greatly improve the classification and the model’s ability to generalize across space and time. Between the two semantic segmentation models, DeepLab provides more accurate horizontal and vertical classifications than FCN when sufficient training data is available. By using DeepLab, the F1 score can be increased by 4 and 10 percentage points for detecting vertical urban growth compared to FCN and RF for Denmark. For mapping the ten other European cities with training data from Denmark, DeepLab also shows an advantage of 6 percentage points over RF for both horizontal and vertical dimensions. The resulting maps across the years 1985 to 2018 reveal different patterns of urban growth between Copenhagen and Aarhus, the two largest cities in Denmark, illustrating that those cities have used various planning policies in addressing population growth and housing supply challenges. In summary, we propose a transferable deep learning approach for automated, long-term mapping of urban form from Landsat images that is effective in areas experiencing a slow pace of urban growth or with small-scale changes. Published article available at: \url{https://doi.org/10.1016/j.rse.2020.112096}

\end{abstract}


\maketitle              
\section{Introduction}

The process of urban densification is a strategy to prevent urban sprawl while addressing housing shortages to accommodate a growing population (Haaland and van Den Bosch, 2015). Many aspects of human well-being, such as air (Stone Jr, 2008), water (\cite{Noorhosseini2017}), and soil quality (\cite{Chuang2018}), ecosystem services (\cite{Seto2012}), and people’s social experiences (\cite{Samuelsson2018}), are related to housing density. Urban planners and scientists can cooperate in designing, developing, and regenerating urban environments based on evidence-based knowledge. Currently, most of the evidence regarding urban densification has accumulated in East Asia, where fast urban growth is occurring (\cite{Chen2020review}). In the meantime, other parts of the world, such as Europe and Australia, are facing slower urban redevelopment and regeneration. The transformation of residential density and green space can take decades to reveal its impacts on human health (\cite{Curtis2004,Engemann2019}). Thus, the long-term observation of slow urban growth is needed for a deeper understanding of sustainable urban design. Using satellite images, the remote sensing community is able to consistently monitor long-term urbanization processes (\cite{Wentz2018}). Myriads of remote sensing studies have mapped impervious areas to illustrate unidirectional urban expansion (\cite{Reba2020}). However, urban growth around the world exhibits a variety of patterns (\cite{Zhu2019}). For instance, while Indian cities appear to horizontally spread out (\cite{Frolking2013,Mahtta2019}), the coastal metropolises in China and South Korea have grown mostly in the vertical dimension, with building height (\cite{Mahtta2019,Zhang2018HV}). Heterogeneity of urban development patterns also appears within countries (\cite{Mahtta2019}). 

To analyze three-dimensional urban growth, scientists have traditionally relied on statistics \cite{He2019,Salvati2013, Zambon2019}). However, official statistics, such as the year of construction and number of building floors, are commonly lacking in spatiotemporal details. For instance, the authorities may only document current buildings (\cite{Zambon2019}) or present aggregated numbers at the block or prefecture level \cite{Salvati2013}). Earth observation can help to solve this lack of data in official statistics, by reconstructing urban morphology in detailed maps. The development of urban morphology remote sensing can be grouped into three types based on data form: LiDAR, high-resolution optical images, and SAR images. Airborne LiDAR is an active sensor that measures ground height. As a product of passive sensors, the optical data itself is not sensitive to ground height, but combing images from different angles (i.e., stereo images) (\cite{Duan2018,Peng2017}) or incorporating shadow detection with high-resolution images (\cite{Shao2011}) can assist building height estimation. Ground-view optical images, such as Google Street View, have also been used to characterize urban morphology in different local climate zones (\cite{Wang2018}) and sky views of street canyons (\cite{Gong2018}). High-resolution Synthetic Aperture Radar (SAR) data, such as TerraSAR-X images, have been used to retrieve building height by radiometric analysis of the typical double-bounce reflection of a building (\cite{Brunner2010,Liao2020}) or by means of advanced SAR interferometric methods such as tomography (\cite{Shi2020,Zhu2010}). However, none of these three forms of data are freely available or consistently recorded across time. Indeed, a recent review quantified the spatiotemporal profiles of urban form detection studies and showed that few studies captured the dynamics of three-dimensional urban growth (\cite{Chen2020review}). The tension between the spatial and temporal data requirements thus exists. In essence, more recent imagery lacks the temporal record, but freely available satellite Landsat imagery before the 1990s has optical signals only at a 30m spatial resolution. 

The United States Geological Survey (USGS) Landsat program has allowed monitoring of land-cover and land-use change since 1972. Landsat images are the most frequently used remote sensing data for tracing urbanization processes (\cite{Deng2018,Song2016,Taubenbock2012}). Based on Landsat 8 images, \cite{Bechtel2015} devised a World Urban Database and Portal Tool (WUDAPT) method to construct a Local Climate Zone (LCZ) map. LCZ is a valuable framework that aims to provide culture-neutral and detailed spatial information for cities in two morphological dimensions: horizontal (compact, open, sparse) and vertical (high, middle, low) (\cite{Stewart2012}). The WUDAPT community commonly uses a non-parametric random forest classifier to deal with varying appearances of Landsat pixels for a LCZ class when conducting large-scale mapping (\cite{Bechtel2019,Demuzere2019,Verdonck2017}). Further progress has been made to better distinguish urban form by employing contextual features. For instance, \cite{Verdonck2017} used moving windows to capture the neighborhood information (e.g., mean, maximum, minimum). Other texture-based RF classifiers have exploited more complex spatial features with co-occurrence matrices (e.g., variance, homogeneity, and contrast) (\cite{Bechtel2016,Xu2017}).  

Deep learning has recently succeeded in making the most of the spatial features in image recognition, classification, and segmentation (\cite{LeCun2015,Zhu2017}). Compared to texture-based machine learning methods, deep learning models do not require hand-crafted features and have shown some success in recent urban form studies (\cite{Rosentreter2020,Liu2020local, Zhang2018ocnn}). Using Landsat images, \cite{Yoo2019} performed the first comparison experiment between CNN and RF approaches to map horizontal and vertical urban form. The study showed the transferability of a CNN classifier when omitting training data in the target city. However, a recent study has warned that the LCZ maps produced from 30m Landsat data might not be satisfactory to reflect the detailed outlines of urban fabric (\cite{Rosentreter2020}). Deep learning approaches and texture-based RFs usually process a group of pixels into one unit (e.g., a patch or a kernel) and thus lose spatial resolution, which results in Landsat-based urban form maps of lowered resolution (90-990 m) (\cite{Bechtel2015,Bechtel2016,Mitraka2015,Zhang2018HV}). Earlier experiments tested the suitability of Landsat data to map horizontal and vertical urban expansion performed by \cite{Zhang2017HV}. The authors also aggregated the binary depictions into 1,000 m cells to obtain “density” information. Nevertheless, Landsat data is the only freely available medium-resolution data we could use to monitor urbanization for the 44 years before the Sentinel era. It is thus critical to develop appropriate approaches for Landsat image analysis to prevent resolution loss while extracting fine-scale patterns of urban form.

\begin{figure*}[t]
    \centering
        \includegraphics[width=0.9\textwidth]{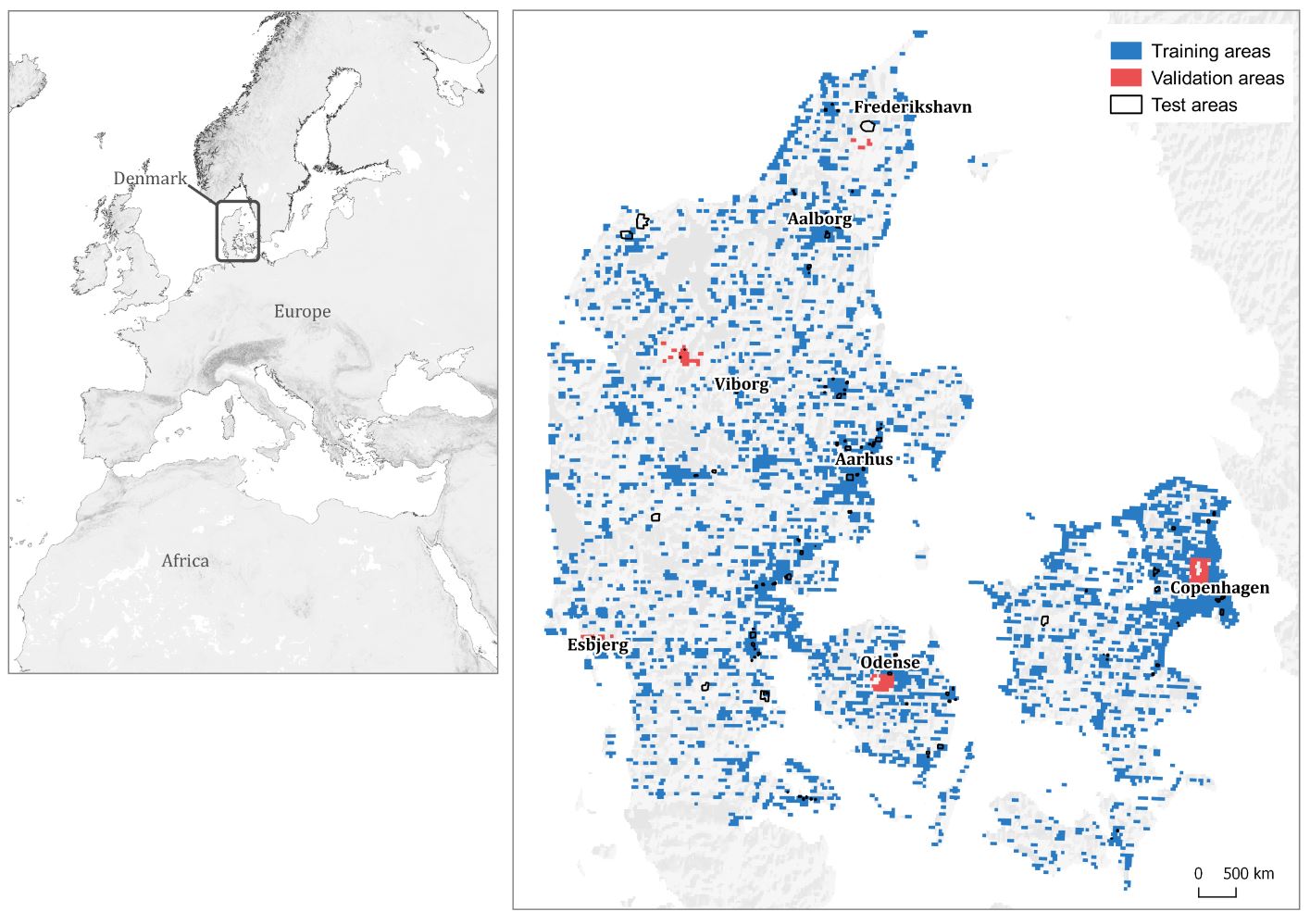}
    \caption{
    Denmark, the study area (training, validation and test areas).
}
    \label{fig:studyarea}
\end{figure*}

Semantic segmentation algorithms facilitated by deep CNNs are increasingly common in the remote sensing field. Unlike scene recognition (i.e., produce one label for one image patch), deep learning models for semantic segmentation aim to provide dense labels in a patch (\cite{Long2015,Chen2017, ,Ronneberger2015}). The advantage of such deep-learning-based semantic segmentation models has attracted the remote sensing community’s attention for delineating the boundaries of trees (\cite{Zabawa2020}), fields (\cite{Waldner2020}), sparse settlements (\cite{Qiu2020fcn}), roads (\cite{Zhang2018road}), and street view objects (\cite{Fang2019}). These applications were primarily made with very high-resolution satellite images, UAV images, point clouds, or at least 20m Sentinel-2 images. We hypothesize that semantic segmentation has a great potential to advance Landsat-based urban mapping because preserving fine spatial resolution is exactly the challenge of Landsat data when using CNN-based methods. We also expect that semantic segmentation using deep networks could capture robust contextual features and provide a more accurate time series analysis of urban dynamics. Yet, to date, multi-annual reference data on vertical urban form are limited or simply not available (\cite{Koziatek2017,Mahtta2019,Sanyal2017,Wurm2013}). This is probably one of the reasons why LCZ studies concentrated on a single slice in time (\cite{Bechtel2016,Verdonck2017, Yoo2019}), despite freely available Landsat time series. Therefore, it is essential to develop models that can generalize to other time points lacking reference data to reconstruct long-term urban records.

In this study, we propose a deep learning semantic segmentation workflow based on Landsat time-series data. It can generate 30m annual maps of urban density in horizontal and vertical dimensions and monitor urban growth. Specifically, our objectives are to: 

\begin{enumerate}
\item compare different strategies for Landsat-based urban from mapping, including texture-based random forest, fully convolutional networks, and DeepLab;
\item assess the spatial and temporal transferability of the models;
\item apply the best-suited model to characterize contemporary urban patterns and urban densification from 1985 to 2018 in Denmark to investigate the relationship between urban form and population growth. 
\end{enumerate}

\begin{figure}[!t]
     \centering
     \resizebox{1\columnwidth}{!}{\includegraphics{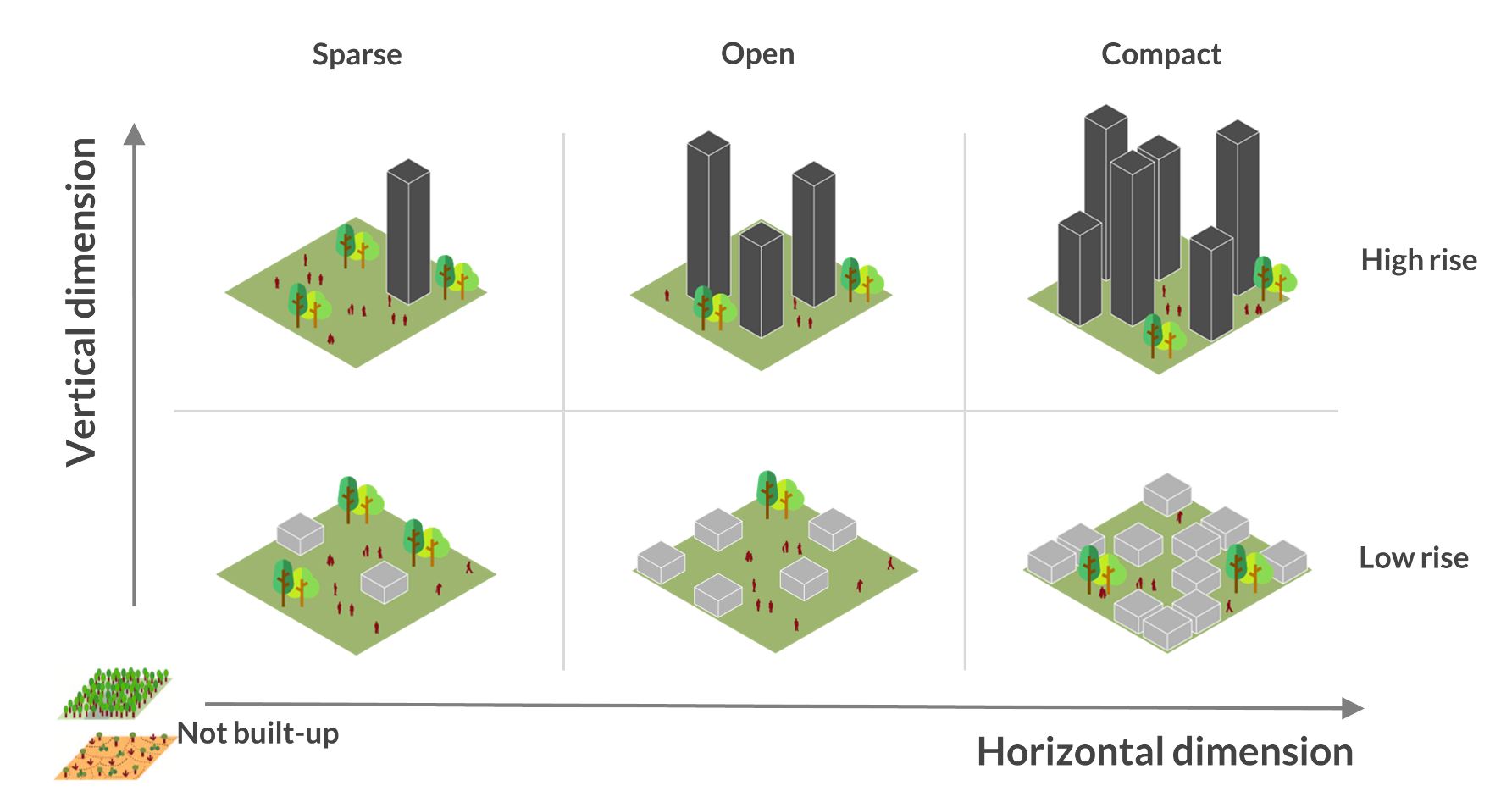}}
     \caption{Classification scheme for urban density in horizontal and vertical dimensions, modified from the local climate zone framework (\cite{Bechtel2015}).}
     \label{fig:LCZ}
\end{figure}

\section{Study area and classification system}

\subsection{Study area}
Denmark is one of the Nordic countries, with a land area of 42,394 km2 (Fig.~\ref{fig:studyarea}). The spatial planning of post-war Denmark is known for its transit-oriented development (TOD) type of development, which aims to maximize the amount of residential, industrial and leisure space within walking distance of public transport. Thus, many new satellite suburbs connected to a central business district (CBD) have emerged alongside the construction of public transport (\cite{Knowles2012}). Despite the shortage of land in the older parts of Danish cities, the majority of population growth has been within the large cities over the last two decades\footnote{BY2: Population 1. January by municipality, size of the city, age and sex. Statistics Denmark. \url{https://www.statbank.dk/BY2}}. Some old neighborhoods have been subject to urban renewal, and their spatial arrangement has been redesigned. For instance, in the single year 2017, approximately four thousand dwellings were newly built in the capital of Denmark, Copenhagen\footnote{Dwellings by region, time, year of construction, use and type of resident. Statistics Denmark. \url{https://www.statbank.dk/tabsel/206367}}, including construction at Copenhagen’s new island Nordhavnen, at the old center Carlsberg town, and the new development along the corridor from the Copenhagen’s CBD to the Danish airway hub Kastrup Airport (\cite{Knowles2012,Lidegaard2018}). 

\subsection{Classification system and reference data}
Our classification scheme to map horizontal and vertical urban density originated from the concept of Local Climate Zones (LCZs) (\cite{Bechtel2015,Stewart2012}). Outlining a classification scheme based on urban morphology, LCZs classify urban landscapes into ten “built” (e.g., compact high, compact low) and seven “natural” zones (e.g., low plants, water). We simplified the LCZ classes by focusing on the built zones and separated them into two dimensions: horizontal and vertical. The horizontal dimension has four classes—compact, open, sparse, not built-up, and the vertical dimension has three classes—high rise, low rise, and not built-up (Fig.~\ref{fig:LCZ}). 

We produced building density labels for each cell at 30m spatial resolution for the years 2014, 2006, and 1995. The horizontal density label was defined by building area ratio, and the vertical density label was produced based on the mean value of building height around the target cell. To label each 30 $\times$ 30m cell, we considered its surrounding 5 $\times$ 5 cells, a block covering an area of 150 $\times$ 150 m (2.25 hectares). The class-defining cell is at the central location of the block. To classify horizontal density, we labeled a cell compact if its block had a building area ratio $>=$0.3, open when it was between 0.15 and 0.3, sparse if it was between 0.02 and 0.15, or not built-up if it was lower than 0.02. To classify vertical density, we labeled a cell high rise if its corresponding block had a building area ratio $>=$0.02 and a mean building height $>=$10 m, low rise if building height \textless10 m, otherwise not built-up.

Our multi-temporal reference data (1995, 2006, 2014) came from the building height datasets generated from nationwide airborne LiDAR surveys for 2006 and 2014 (Angelidis et al., 2017), the digitized building boundaries of Denmark, and national aerial photos for 1995, 2006, and 2014. The LiDAR surveys for 2006 and 2014 provided the digital surface models (DSMs) with height information above the ground. The LiDAR data has a horizontal accuracy of 0.6 m and 0.15m, and vertical accuracy of 0.07 m and 0.06 m for 2006 and 2014, respectively. We used data for entire Denmark for 2006 and 2014 to produce urban density labels throughout our study area. The aerial photos for 1995, 2006, and 2014 were visually analyzed to obtain the 1995 labels and to correct the 2006 and 2014 labels in the test areas.

\begin{figure*}[!t]
    \centering
        \includegraphics[width=0.9\textwidth]{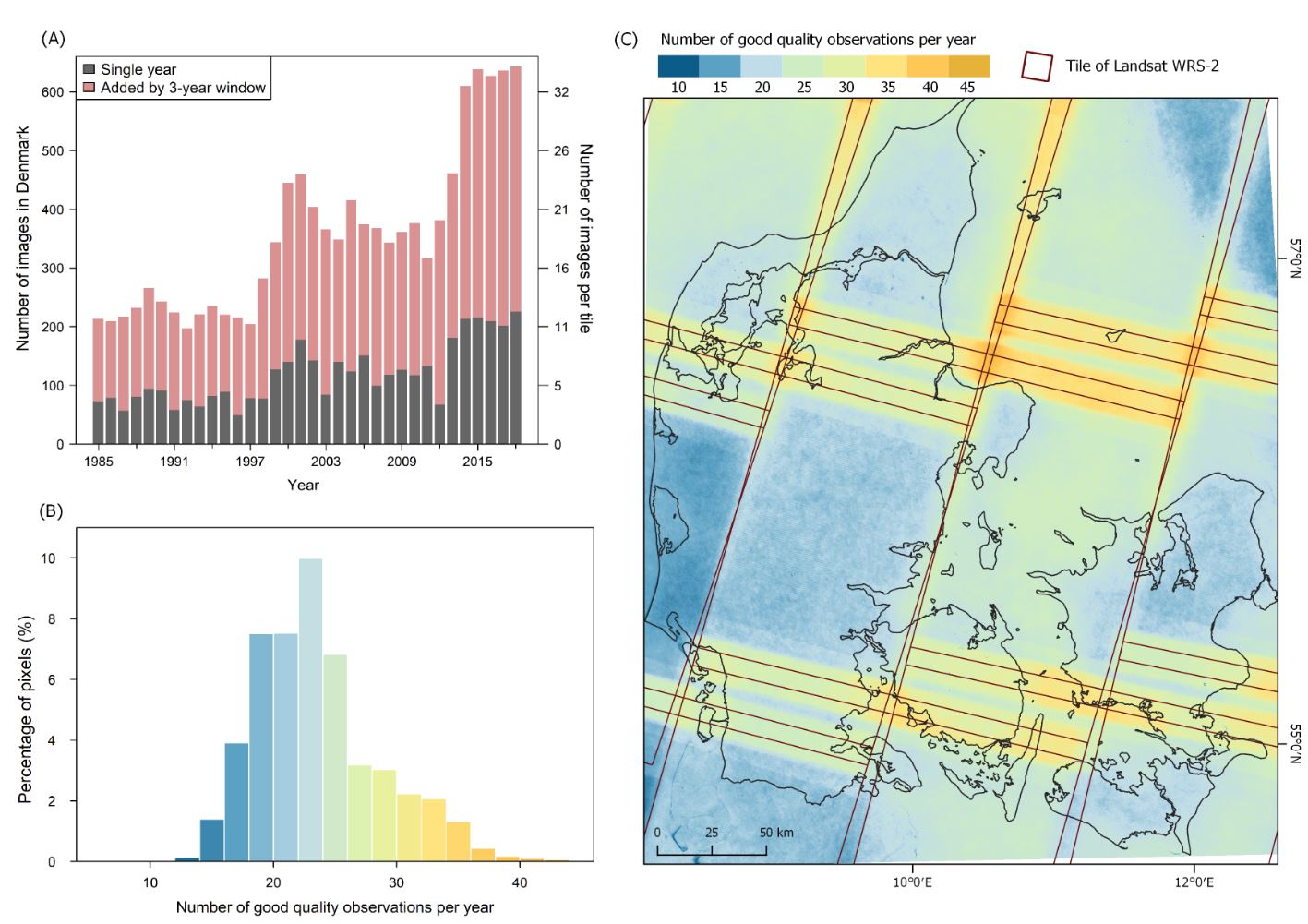}
    \caption{Availability of time series Landsat images from 1985 to 2018. (A) The number of images by sensors Landsat 5/7/8 per single year and the 3-year rolling window used in this study, (B) the percentage distribution of good quality observations of the pixels in the study area, and (C) the spatial distribution of the number of good quality observations per year and Landsat Worldwide Reference System 2 (WRS-2) tiles in Denmark.}
    \label{fig:images}
\end{figure*}

\section{Methods} 
The approach entailed four steps to map urban form on an annual basis (Fig.~\ref{fig:methods}). First, we acquired and pre-processed cloud-free annual Landsat data from 1985 to 2018 from Google Earth Engine (Section~\ref{chap:Image}), and ensured the consistency of the time-series images (Section~\ref{chap:Timeseries}). We separated the whole dataset for the training year into training and validation datasets, and used other years for test datasets (Section~\ref{chap:Training}). Second, we trained our CNN-based semantic segmentation models and a widely-used random forest classifier to map urban density in horizontal and vertical dimensions (Section~\ref{chap:Models}). Third, we statistically compared the performance of the three approaches. We also compared the transferability of the three models for mapping to another year for Denmark and across ten non-Danish cities for 2015 when omitting training data (Section~\ref{chap:Assessment}). Finally, we used the validated model to map annual urban form for Denmark explicitly from 1985 to 2018. We demonstrate how these maps can be used to identify various patterns of urban growth together with population data at regional and national scales (Section~\ref{chap:Characterizing}).

\subsection{Image collection} \label{chap:Image}
We pre-processed images through Google Earth EngineTM (\url{https://earthengine.google.com/}). Google Earth Engine provides online access to USGS Landsat data (Huang et al., 2017). We used all available Level-2 Landsat-5, Landsat-7, and Landsat-8 images to create annual composites from 1985 to 2018 for the Landsat tiles covering Denmark (Fig.~\ref{fig:images}C). In total, those images span 19 path/row tiles of the Landsat worldwide reference system (WRS-2). The number of images per year in Denmark ranges from 49 in 1996 to 225 in 2018 (Fig.~\ref{fig:images}A). Level-2 products are surface reflectance data at a 30m spatial resolution with systematic atmospheric correction and are provided with per-pixel quality information assessed with the Fmask algorithm (\cite{Zhu2015}). We used only good quality observations where pixels covered by clouds, cloud shadows with a high confidence level, and scan-line corrector (SLC)-off gaps from Landsat-7 were removed. Only 0.01\% of the pixels in the study area have less than ten good quality observations per year (Fig.~\ref{fig:images}B, C). 

\subsection{Time-series processing} \label{chap:Timeseries}
Time-series analysis of urban density can be affected by signal noise. We utilized four strategies to improve the consistency of time series data quality: (1) time filtering, (2) three-year rolling window, (3) standardization, and (4) post-classification smoothing of the annual maps.

\subsubsection{Time filtering} 
The environment surrounding human settlements is known to be affected by seasonal variations in vegetation, such as agricultural activities and gardens. Bright snow cover also causes significant interference in the detection of the built environment. To avoid snow and retain high-level vegetation signals (e.g., near-infrared), we only used images from May to August each year, which helps focus on long-term urban development instead of intra-annual variations.

\subsubsection{Three-year moving window} 
To produce cloud-free annual composites, we used a three-year moving window to minimize atmospheric noise and inter-annual phenological variations (\cite{Song2016}). The number of observable pixels per year also increased by using the three-year window (Fig.~\ref{fig:images}). We chose the median of each reflective band because it can capture abrupt changes occurring in the middle year (\cite{Sagar2017,Song2016}). 

\subsubsection{Standardization} 

Surface reflectance values were standardized and normalized to remove environmental noise remaining after atmospheric correction (\cite{Sexton2013}), e.g., the brightening effect of thin clouds or sun glint close to water. The bands of each pixel and for each year were then standardized to a joint range using:
\vspace{5mm}

$\rho_{ib}’ = \frac{\rho_{ib}}{max(\rho_b)}$
\vspace{5mm}

where $\rho_{ib}$ is the reflectance of band $b$ in pixel $i$, standardised by dividing by $max(\rho_b)$, which was defined as the 99.5 percentile value in the training year, i.e., 2014, in that band: blue (0.1024), green (0.1374), red (0.1532), near-infrared (0.4679), shortwave-infrared 1 (0.2872), and shortwave-infrared 2 (0.2207). Consistent maximum values applied to the annual time-series can preserve differences between years.

\subsubsection{Post-classification smoothing} 
After performing the classifications, we applied a Savitzky-Golay filter to remove abnormal observations across the classification time series. This filter has been used in previous urban remote sensing studies (\cite{Li2018SG}). The Savitzky-Golay filter is also known for detecting complex behavior phenological time-series (\cite{Joensson2004}). We hypothesized that this method could capture a rapid decrease in urban density, such as when removing old houses.

\subsection{Training, validation, and test data} \label{chap:Training}
We addressed three issues: (1) data redundancy, (2) class imbalance, and (3) patch heterogeneity when preparing training data. First, CNN-based segmentation models require a training data that is large but not redundant. To deal with this issue, we defined a minimum distance of 150 m between the labeled cells to reduce spatially correlated data. Second, built-up areas often represent an insignificant share compared to not-built-up areas. Therefore, we reduced the sample size of the dominant class (\textit{not built-up}) to be at most five times the second frequent class. Third, heterogeneity within patches is essential to ensure that the segmentation model can precisely delineate the boundaries between classes. To increase the proportion of heterogeneous patches, we removed the training patches that contain only \textit{not built-up} labels. 

All data for the year 2014 were split into training and validation sets. The semantic segmentation models were developed based on the training data (N of patches = 26,413). The validation set is needed to estimate the model performance in each training iteration. We used the validation set for the same year 2014 (N of patches=686) to diagnose overfitting, which was collected from the sites that were not used for training. The best model trained and validated on 2014 data was used to map urban density from 1985 to 2018. 

For accuracy assessment, we selected several random test sites (N of pixels = 9,900) and extracted data for the years 1995, 2006, and 2014. We reported the final accuracy only based on the test data for 1995 and 2006. The test data for 2014, which is overlapping with the training data, was only used for comparison – whether the classifications of other years (i.e., 1995 and 2006) can reach accuracies as high as the training year (i.e., 2014). The geography of the training, validation, and test sites and the class proportions are presented in Fig.~\ref{fig:studyarea} and Fig. S1-2, Appendix A.  

We derived a test dataset on positive categorical change (urban growth) by comparing the reference labels for the test sites for 2006 and 2014. We labeled a pixel \textit{growth} if it was higher or denser in 2014 compared to 2006 (e.g., \textit{not built-up} or \textit{low-rise} in 2006 and \textit{high-rise} in 2014), otherwise \textit{no growth}. We compared the aerial photos of these two years to double-check if the change occurred and corrected labels of false growth if the buildings looked the same between the photos. The derived test data on change consists of bi-temporal growth in the horizontal dimension (N = 905) and vertical dimension (N = 764).

Additionally, we used an open-access dataset of LCZ for testing spatial transferability of the models. The LCZ dataset comprises polygons, which were visually interpreted around 2015 by the WUDAPT community (\cite{Bechtel2019}). The datasets cover ten European cities, which are Wageningen, Arnhem (the Netherlands), Berlin, Hamburg, Augsburg (Germany), Brussels, Antwerp, Leuven (Belgium), and Paris, Nantes (France) (Fig. S3-4, Appendix A). We used stratified random sampling for the transferability test. First, we converted polygons of the 17 LCZ classes into three vertical and four horizontal classes while excluding water and heavy industry classes in LCZs due to its mixed form (detailed in Table S1, Appendix A). We used the labeled polygons as our stratified test sites. Second, we randomly sampled 200 points from those polygons for each horizontal built-up class and 600 points for the \textit{not built-up} class (same amount as the built-up classes in total) per city. We collected 300 points for each vertical built-up class and 600 points for the \textit{not built-up} class per city. A minimum distance of 31 m was set to ensure no cells overweight. Due to the restriction, the datasets for a few cities did not reach the target sample size (Table S2, Appendix A). In sum, a sample of 10,851 points was derived for the horizontal dimension and 11,295 points for the vertical axis across the ten non-Danish-cities to assess spatial transferability.

\begin{figure*}[!t]
    \centering
        \includegraphics[width=0.9\textwidth]{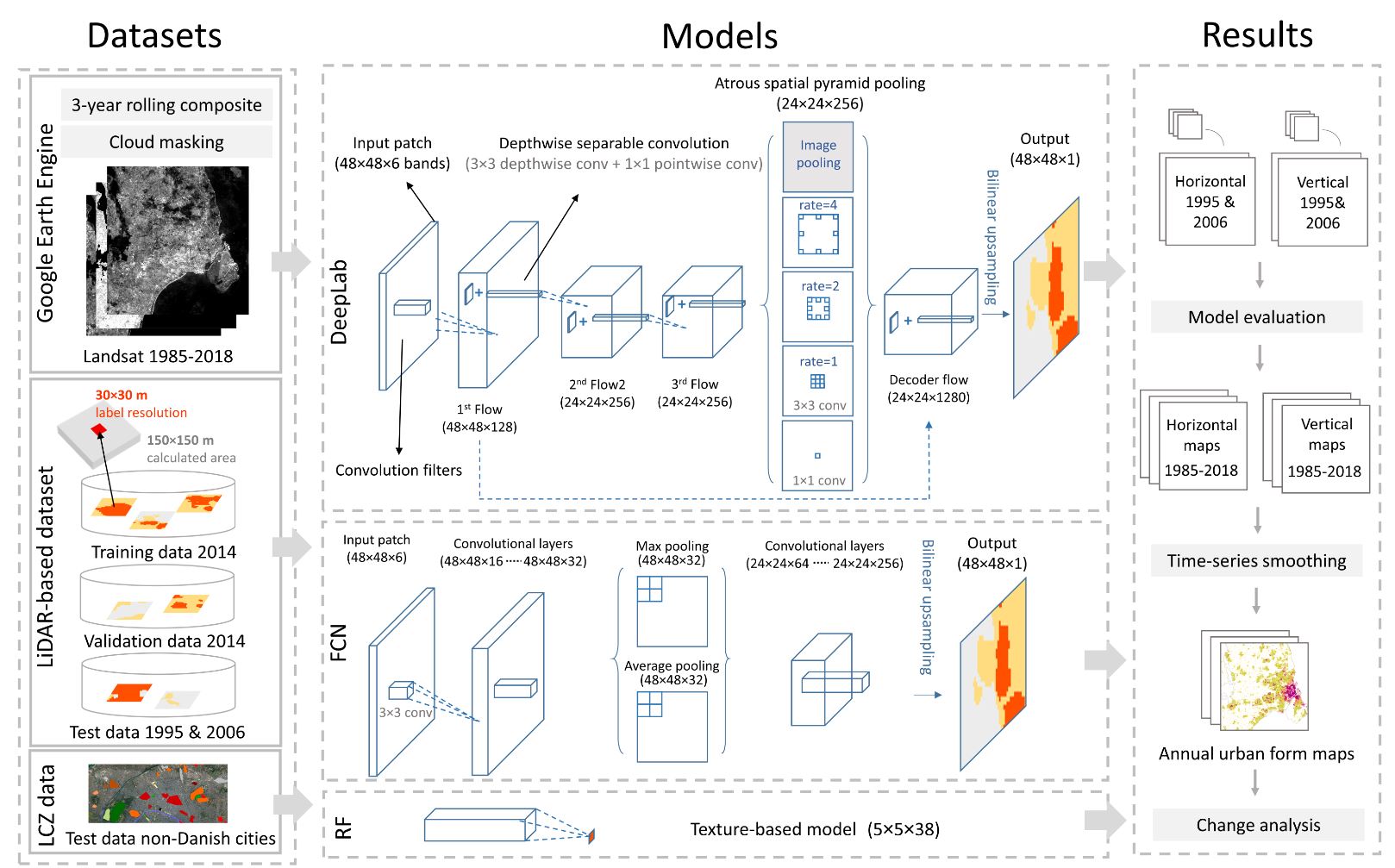}
    \caption{Structure of the proposed approach.
}
    \label{fig:methods}
\end{figure*}

\subsection{Models} \label{chap:Models}
\subsubsection{Random forest} 
We used texture-based random forest (RF) as our benchmark model to test our hypothesis that our proposed deep learning models can provide higher accuracy than a non-deep-learning model. RF has shown success in model generalization (\cite{Belgiu2016}) and urban form mapping, among other classic machine learning methods (\cite{Bechtel2016}). We used 38 layers, including 30 spectral attributes (maximum, minimum, mean, median, and standard deviation of all the six bands), and 8 texture features from gray-level co-occurrence matrix (GLCM) based on the first principal component of the bands. The texture features are composed of mean, variance, homogeneity, contrast, dissimilarity, entropy, second moment, and correlation (\cite{Xu2017}). 

\subsubsection{Semantic segmentation models: DeepLab and FCN} 

Semantic segmentation frameworks aim to assign labels to each pixel within an image patch by an encoder-decoder structure (\cite{Long2015}). A state-of-the-art semantic segmentation architecture, DeepLab, has achieved success in numerous benchmark datasets such as the Cityscapes Dataset, which covers a variety of urban landscapes (\cite{Chen2018Atrous}). To extract rich multi-scale contextual information while avoiding downsampling, it utilizes atrous spatial pyramid pooling (ASPP) (\cite{Chen2017}). The atrous rates determine the scales used for contextual feature extraction, which should adapt to Landsat data for urban density mapping. 

\begin{figure*}[!b]
    \centering
        \includegraphics[width=0.9\textwidth]{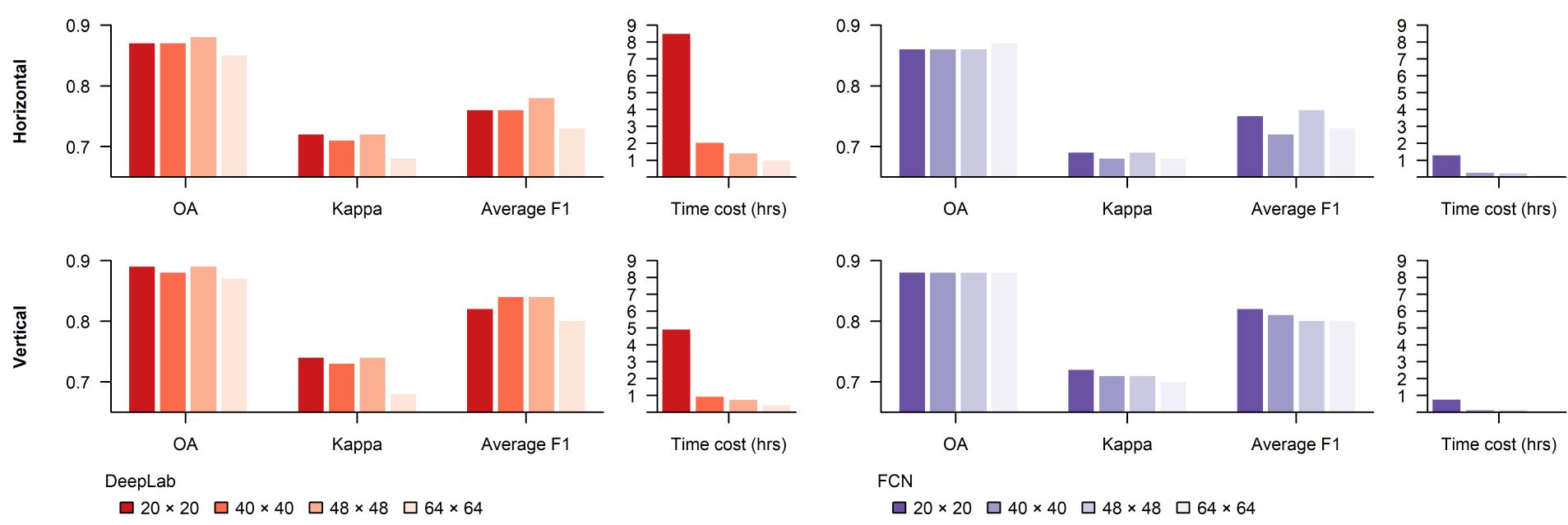}
    \caption{Overall accuracy (OA), Kappa coefficient, average F1 score, and time cost resulting from different patch sizes and semantic segmentation approaches.}
    \label{fig:patchsize}
\end{figure*}

To construct a backbone of our DeepLab network (Fig.~\ref{fig:methods}), we used the Xception architecture (\cite{Chollet2017}). Xception is characterized by cost-effective depthwise separable convolutions to deal with large scale mapping. We utilized a stride of one for the convolutions in the first two blocks in the first flow, which led to an unchanged patch size to avoid resolution loss from the Landsat input images. In the third block of the first flow, we used a stride of two, leading to a patch size of half the input size (H/2, W/2). We kept a stride of one in the blocks of the second and the third flows, which retained the patch size (H/2, W/2) (Fig.~\ref{fig:methods}). To effectively capture multi-scale contextual features, we utilized atrous pyramid pooling with one 1 $\times$ 1 convolution, three 3 $\times$ 3 convolutions, and the image-level feature to represent the spatial features from local to global levels (Fig.~\ref{fig:methods}). The output features of all five branches in atrous pyramid pooling were concatenated into a decoder flow and passed through another 1 $\times$ 1 convolution. To support the decoding, the feature maps were concatenated with the low-level feature from the first flow and subsequently used for the final prediction with two 3 $\times$ 3 convolutions and one 1 $\times$ 1 convolution. Finally, a bilinear upsampling layer resulted in an output with a shape the same as the input (H, W).

In addition, we investigated the performance of a simple fully convolutional network (hereafter FCN), which has shown promising results in human settlement mapping (Long et al., 2015; Qiu et al., 2020b). The proposed FCN architecture is illustrated in Fig.~\ref{fig:methods}. In short, its encoder does not downsample as much, and its decoder uses bilinear interpolation to localize the predictions. The encoder consists four convolutional layers to extract low-level features from original inputs of a size (H, W), maximum pooling and average pooling layers to downsample the patches into (H/2, W/2), four following convolutional layers to capture high-level features from the downsampled patch (H/2, W/2), and the last convolutional layer for prediction. The prior eight convolutional layers use 3 $\times$ 3 convolutions and the final convolutional layer uses 1 $\times$ 1 convolution.

\subsubsection{Experimental setting of DeepLab, FCN, and RF} 
We adapted patch size (DeepLab and FCN) and atrous rates (DeepLab), which may affect the performance of semantic segmentation models. The influence of patch size was investigated among 20 $\times$ 20, 40 $\times$ 40, 48 $\times$ 48, and 64 $\times$ 64 candidates. The atrous rates of convolutions for DeepLab, which affect the scales of feature extraction, were tested from (1, 2, 3), (1, 2, 4), (1, 2, 6), and (2, 4, 8), ranging from local to more global features. We tuned other deep learning parameters empirically. Specifically, the learning rate was chosen as 0.0002, the step to create patches was set to 24 (image patches overlapping by 50\%) to avoid edge effect, and the number of epochs was set as 12 to learn features fully. The best-trained model among the 12 epochs was selected based on validation data to trade off bias and variance. To ensure a fair comparison for the RF model, we examined the window size for extracting texture features from 3 $\times$ 3, 5 $\times$ 5, 7 $\times$ 7, and 47 $\times$ 47. We used 200 trees after a sensitivity analysis among 25, 50, 100, 200, 500, and 1000 trees. The adapted models were used to predict each pixel across Denmark.

\subsection{Assessment of accuracy and transferability} \label{chap:Assessment} 

We used the test data for 1995 and 2006 to evaluate the models trained on the 2014 dataset. The metrics adopted include overall accuracy, kappa coefficient, user’s accuracy, producer’s accuracy, F1 score, and average F1 score. The user’s accuracy represents the model performance in reducing overestimation, and the producer’s accuracy refers to the ability to minimize the underestimation (\cite{Olofsson2014}). The average F1 score is the average harmonic mean of the user’s and producer’s accuracy of all thematic classes regardless of their sample size. We implemented a non-parametric McNemar’s test (\cite{Foody2004}) to assess whether the deep learning approaches are statistically more accurate than a RF classifier with texture features.  McNemar’s test allowed us to compare models without resampling, which helped to evaluate large and slow-to-train deep learning models.

We tested the models’ temporal transferability based on multi-annual reference data and the growth labels (Section~\ref{chap:Training}). First, we evaluated the accuracy for different years, comprising 2014 (with training data), 2006 and 1995 (without training data). Second, we tested the accuracy of urban growth in low-density (i.e., not built-up became sparse built) and high-density developed areas (i.e., sparse or not built-up became open or compact built). This can indicate the applicability of our proposed model to change analysis. 
We also investigated the sensitivity to sample size by dropping out training data iteratively from 10\%, 30\%, 50\%, 70\% to 80\% with 20-fold resampling for each training data size. As such, the mean and standard deviation of accuracy were reported.

We validated the models’ spatial transferability to European cities while the model was trained exclusively on Danish data. We applied our trained models to these cities using the Landsat images in 2015 (matched with the reference dataset) with the same pre-processing approach as used for Denmark to derive horizontal and vertical urban form maps (Fig. S3-4, Appendix A). Then, we used the crowdsourced test dataset (Section~\ref{chap:Training}) to estimate each city’s accuracy metrics.

\begin{figure*}[!b]
    \centering
        \includegraphics[width=0.8\textwidth]{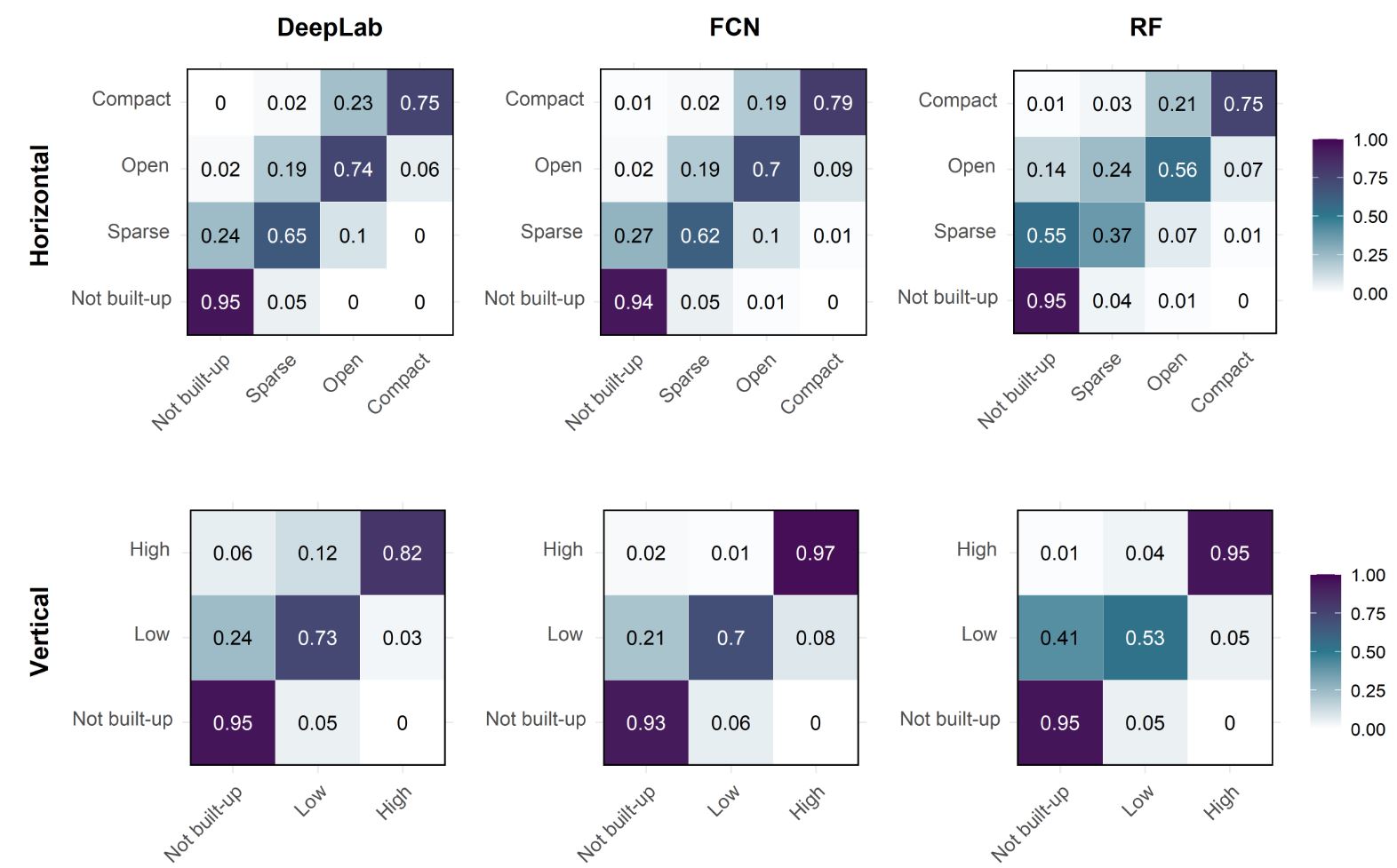}
    \caption{Confusion matrices (x-axis: reference, y-axis: map) of urban density classification in horizontal and vertical dimensions using different methods.}
    \label{fig:matrix}
\end{figure*}

\subsection{Characterizing changes in urban density} 
\label{chap:Characterizing} 

Based on the best deep learning model, we derived the urban density maps in horizontal and vertical dimensions from 1985 to 2018. We combined the two dimensions into annual urban form maps, consisting of classes \textit{compact high, compact low, open high, open low, sparse high, sparse low}, and \textit{not built-up} classes. We measured the annual area of each urban form at national and regional scales. To characterize the different types of urban development, we observed changing urban form in three types of cities: the capital, capital suburbs, and provincial cities. We calculated the magnitude of changes in vertical and horizontal dimensions for the study period. We dropped the beginning and the end of the time series to perform smoothing. For the remaining period from 1987 to 2017, we coupled population growth trends from Statistics Denmark data (\url{https://www.statistikbanken.dk/}) with trends of urban density in the two dimensions to investigate their relationship.

\section{Results}
\label{sec:Results}

\subsection{Experimental results: DeepLab, FCN, and RF}
The training of DeepLab took approximately 1.4 hours, and FCN took 14 minutes for entire Denmark (42,394 km2), which ran on a server featuring a TITAN X GPU with 3584 cores, while the texture-based RF took about 14 minutes using 12 CPUs in parallel. Atrous rates of 1, 2, and 4 produced the highest kappa and average F1 score, meaning that the contextual information at these scales combined was useful to boost accuracies for the thematic urban form classes (Table~\ref{tab:atrous}). The DeepLab model reached the highest accuracy with 48 $\times$ 48 patch size (Fig.~\ref{fig:patchsize}). The smaller the patch size, the longer the training took. However, when using varying patch size to train FCN, the model produced very similar overall accuracy (Fig.~\ref{fig:patchsize}). The experimental results on RF showed that 5 $\times$ 5 was the optimal window size for extracting texture information for the urban form prediction tasks, while the large window size 47 $\times$ 47 led to poor score (Table S3, Appendix A).

\vspace{0.1cm}
\begin{table}[H]
\caption{Results of DeepLab with different atrous rates. OA: overall accuracy.}
\centering
\setlength{\tabcolsep}{5pt}
\centering
\label{tab:atrous}
\resizebox{\columnwidth}{!}{%
\begin{tabular}{lllllll}
\hline
\multirow{2}{*}{Atrous rate} & \multicolumn{3}{l}{Horizontal   density model} & \multicolumn{3}{l}{Vertical density model} \\ \cline{2-7} 
                             & OA            & Kappa       & Ave. F1       & OA          & Kappa      & Ave. F1      \\ \hline
1, 2, 3                      & 87.2\%        & 0.71        & 0.76             & 88.6\%      & 0.72       & 0.82            \\
\textbf{1, 2, 4}                      & 87.7\%        & 0.72        & 0.78             & 88.9\%      & 0.74       & 0.84            \\
1, 2, 6                      & 86.4\%        & 0.69        & 0.71             & 88.2\%      & 0.71       & 0.82            \\
2, 4, 8                      & 87.4\%        & 0.71        & 0.75             & 87.2\%      & 0.70       & 0.80            \\ \hline
\end{tabular}}
\end{table}

In general, the semantic segmentation models, DeepLab and FCN, were more accurate than the RF classifier (Fig.~\ref{fig:matrix}). For urban density maps in horizontal and vertical dimensions, DeepLab and FCN achieved an overall accuracy of 86-88\%; both models were statistically more accurate than texture-based RF (below 81\%) (Table~\ref{tab:accuracy}). 

\begin{figure}
    \centering
    \resizebox{1\columnwidth}{!}{\includegraphics{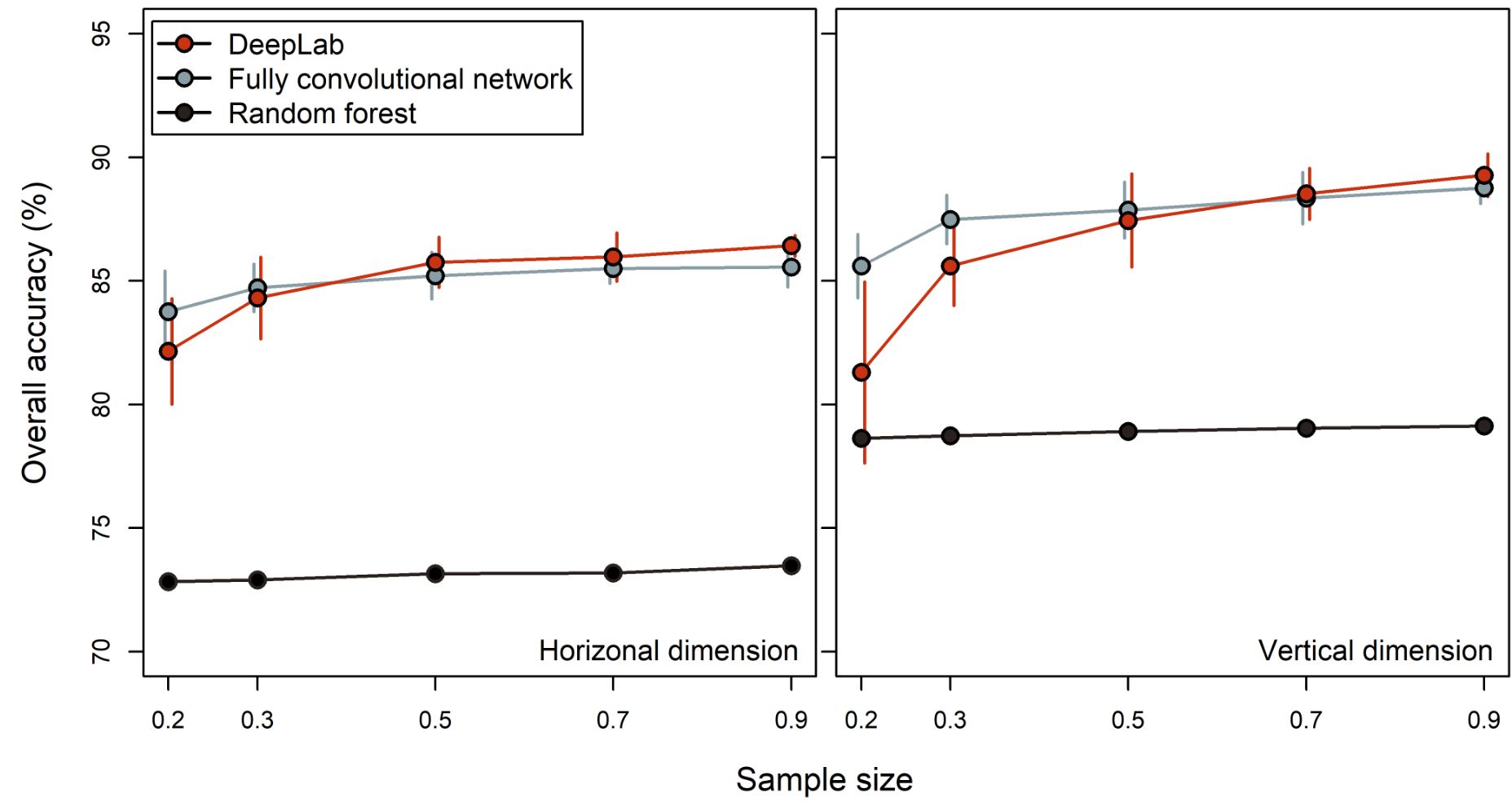}}
    \caption{Sensitivity analysis of sample size (90\%, 70\%, 50\%, 30\%, and 20\% of the original training sample size). The segments show the standard deviation of accuracies from 20-fold resampling runs.}
    \label{fig:samplesize}
\end{figure}

\begin{figure*}[!t]
    \centering
        \includegraphics[width=0.9\textwidth]{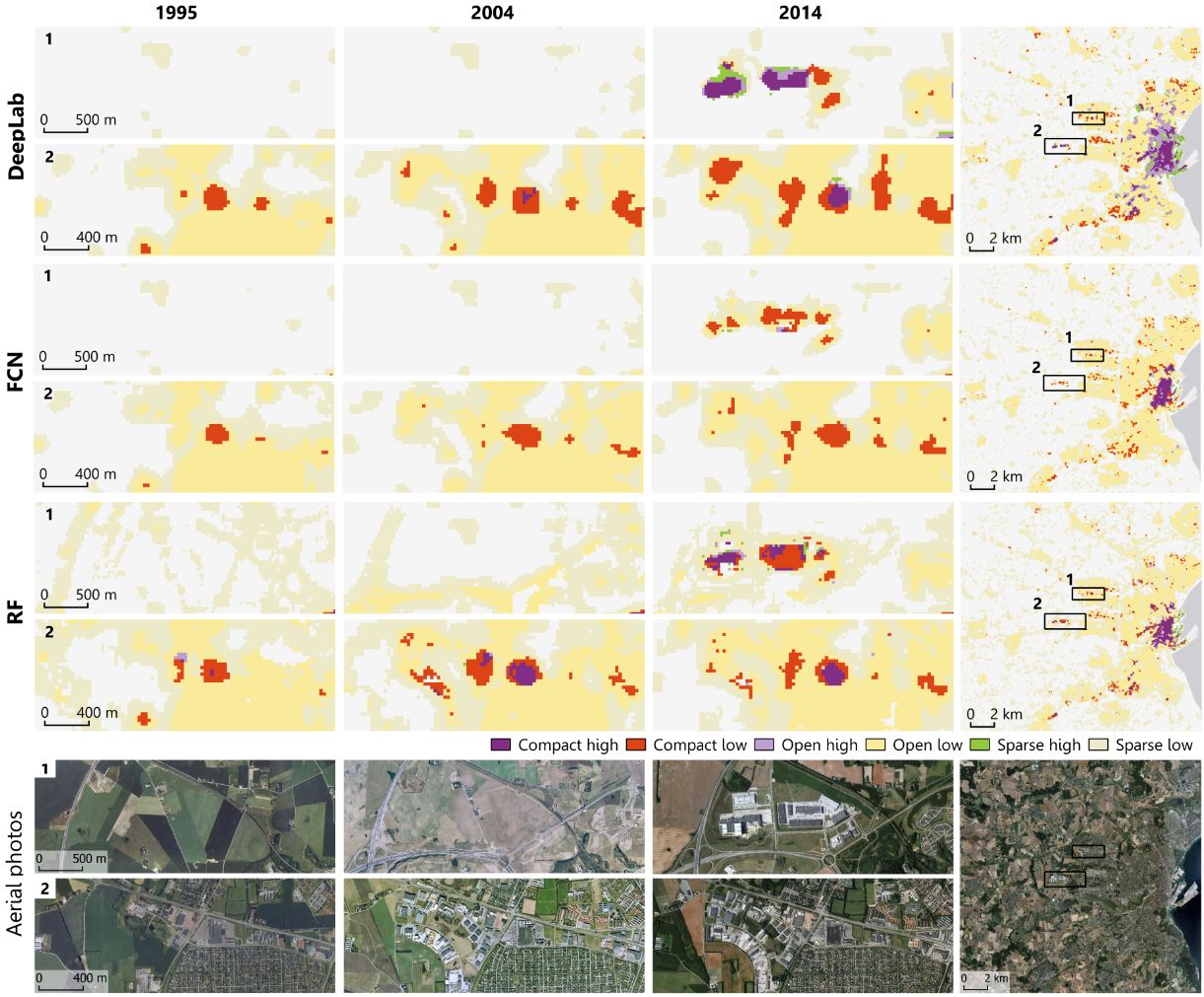}
    \caption{Examples of urban form dynamics produced by DeepLab, FCN, and texture-based RF.}
    \label{fig:timecompare}
\end{figure*}

DeepLab led to the most balanced producer’s and user’s accuracies among the thematic classes (Table~\ref{tab:accuracy}). FCN reached similar scores for horizontal density classification, while unbalanced classification results were observed for vertical density mapping. For instance, FCN struggled to detect high-rise buildings when surrounding areas were open (e.g., large shopping malls in the suburbs, Fig.~\ref{fig:timecompare}). The RF classifier notably underperformed for the ‘medium density’ classes. Besides, the RF was prone to misclassify motorways and sparse settlements, showing its difficulty in distinguishing impervious surfaces with and without buildings (Fig.~\ref{fig:timecompare}). In comparison, the DeepLab model had higher user’s accuracies for \textit{compact} and \textit{open} classes than the \textit{sparse} class. The confusion was more pronounced between the \textit{sparse} and the \textit{not built-up} classes, such as tilled agricultural fields nearby farmhouses (Fig.~\ref{fig:timecompare}). For vertical density yielded by DeepLab, the user’s accuracy for the high rise (82\%) was higher than the \textit{low rise} (73\%).

\begin{table*}[!t]
\caption{Accuracies and 95\% confidence intervals for horizontal and vertical thematic classes based on the test data for 1995 and 2006.}
\centering
\setlength{\tabcolsep}{5pt}
\centering
\label{tab:accuracy}
\resizebox{0.73\textwidth}{!}{%
\begin{tabular}{lllllllclclllclclllrlrl}
\hline
\multicolumn{3}{l}{Horizontal model}            & \multicolumn{3}{l}{Overall}                    &                      & \multicolumn{5}{l}{User's}                                 &                      & \multicolumn{5}{l}{Producer's}                             &                      & \multicolumn{4}{l}{McNemar's   test}                            \\ \cline{4-6} \cline{8-12} \cline{14-18} \cline{20-23} 
                     & \multicolumn{2}{l}{}     & accuracy         & \multicolumn{2}{r}{95\% CI} &                      & \multicolumn{2}{l}{accuracy} & \multicolumn{3}{c}{95\% CI} &                      & \multicolumn{2}{l}{accuracy} & \multicolumn{3}{c}{95\% CI} &                      & \multicolumn{2}{r}{Chi2} & \multicolumn{2}{l}{p-value}          \\ \hline
\multicolumn{4}{l}{Random   forest}                                & \multicolumn{2}{l}{}        &                      & \multicolumn{2}{l}{}         & \multicolumn{3}{l}{}        &                      & \multicolumn{2}{l}{}         & \multicolumn{3}{l}{}        &                      & \multicolumn{2}{l}{}     & \multicolumn{2}{l}{}                 \\
                     & Compact          & \multicolumn{2}{l}{}     & \multicolumn{2}{l}{}        &                      & \multicolumn{2}{c}{75}       & \multicolumn{3}{c}{±2.9}    &                      & \multicolumn{2}{c}{75.6}     & \multicolumn{3}{c}{±2.8}    &                      & \multicolumn{2}{r}{—}    & \multicolumn{2}{l}{}                 \\
                     & Open             & \multicolumn{2}{l}{}     & \multicolumn{2}{l}{}        &                      & \multicolumn{2}{c}{55.7}     & \multicolumn{3}{c}{±2.5}    &                      & \multicolumn{2}{c}{72.6}     & \multicolumn{3}{c}{±2.5}    &                      & \multicolumn{2}{r}{—}    & \multicolumn{2}{l}{}                 \\
                     & Sparse           & \multicolumn{2}{l}{}     & \multicolumn{2}{l}{}        &                      & \multicolumn{2}{c}{37.3}     & \multicolumn{3}{c}{±1.2}    &                      & \multicolumn{2}{c}{56.3}     & \multicolumn{3}{c}{±1.6}    &                      & \multicolumn{2}{r}{—}    & \multicolumn{2}{l}{}                 \\
                     & Not   built-up   & \multicolumn{2}{r}{73.8} & \multicolumn{2}{c}{±0.5}    &                      & \multicolumn{2}{c}{95.4}     & \multicolumn{3}{c}{±0.4}    &                      & \multicolumn{2}{c}{80.2}     & \multicolumn{3}{c}{±0.7}    &                      & \multicolumn{2}{r}{—}    & \multicolumn{2}{l}{}                 \\
\multicolumn{5}{l}{Fully convolutional network}                                   & \multicolumn{3}{l}{}                                & \multicolumn{2}{l}{}    &         & \multicolumn{3}{l}{}                          & \multicolumn{2}{l}{}    &         & \multicolumn{3}{l}{}                        & \multicolumn{2}{l}{}            &                 \\
                     & Compact          & \multicolumn{2}{l}{}     & \multicolumn{2}{l}{}        &                      & \multicolumn{2}{c}{78.6}     & \multicolumn{3}{c}{±2.9}    &                      & \multicolumn{2}{c}{79.3}     & \multicolumn{3}{c}{±2.9}    &                      & \multicolumn{2}{r}{4}    & \multicolumn{2}{r}{0.05}             \\
                     & Open             & \multicolumn{2}{l}{}     & \multicolumn{2}{l}{}        &                      & \multicolumn{2}{c}{70.4}     & \multicolumn{3}{c}{±2.4}    &                      & \multicolumn{2}{c}{62.4}     & \multicolumn{3}{c}{±2.4}    &                      & \multicolumn{2}{r}{132}  & \multicolumn{2}{r}{\textless{}0.001} \\
                     & Sparse           & \multicolumn{2}{l}{}     & \multicolumn{2}{l}{}        &                      & \multicolumn{2}{c}{61.9}     & \multicolumn{3}{c}{±1.6}    &                      & \multicolumn{2}{c}{67.8}     & \multicolumn{3}{c}{±1.7}    &                      & \multicolumn{2}{r}{1322} & \multicolumn{2}{r}{\textless{}0.001} \\
\multicolumn{1}{r}{} & Not   built-up   & \multicolumn{2}{r}{86.4} & \multicolumn{2}{c}{±0.4}    &                      & \multicolumn{2}{c}{94.2}     & \multicolumn{3}{c}{±0.4}    & \multicolumn{1}{r}{} & \multicolumn{2}{c}{93.4}     & \multicolumn{3}{c}{±0.4}    & \multicolumn{1}{r}{} & \multicolumn{2}{r}{1360} & \multicolumn{2}{r}{\textless{}0.001} \\
\multicolumn{2}{l}{DeepLab}             & \multicolumn{2}{l}{}     & \multicolumn{2}{c}{}        &                      & \multicolumn{2}{c}{}         & \multicolumn{3}{c}{}        & \multicolumn{1}{r}{} & \multicolumn{2}{c}{}         & \multicolumn{3}{c}{}        & \multicolumn{1}{r}{} & \multicolumn{2}{r}{}     & \multicolumn{2}{r}{}                 \\
\multicolumn{1}{r}{} & Compact          & \multicolumn{2}{l}{}     & \multicolumn{2}{c}{}        &                      & \multicolumn{2}{c}{74.5}     & \multicolumn{3}{c}{±2.8}    & \multicolumn{1}{r}{} & \multicolumn{2}{c}{88.8}     & \multicolumn{3}{c}{±2.2}    & \multicolumn{1}{r}{} & \multicolumn{2}{r}{6}    & \multicolumn{2}{r}{0.02}             \\
\multicolumn{1}{r}{} & Open             & \multicolumn{2}{l}{}     & \multicolumn{2}{c}{}        &                      & \multicolumn{2}{c}{74.4}     & \multicolumn{3}{c}{±2.4}    & \multicolumn{1}{r}{} & \multicolumn{2}{c}{63.3}     & \multicolumn{3}{c}{±2.4}    & \multicolumn{1}{r}{} & \multicolumn{2}{r}{196}  & \multicolumn{2}{r}{\textless{}0.001} \\
\multicolumn{1}{r}{} & Sparse           & \multicolumn{2}{l}{}     & \multicolumn{2}{c}{}        &                      & \multicolumn{2}{c}{65.5}     & \multicolumn{3}{c}{±1.6}    & \multicolumn{1}{r}{} & \multicolumn{2}{c}{68.1}     & \multicolumn{3}{c}{±1.6}    & \multicolumn{1}{r}{} & \multicolumn{2}{r}{1519} & \multicolumn{2}{r}{\textless{}0.001} \\
                     & Not built-up     & \multicolumn{2}{r}{87.7} & \multicolumn{2}{c}{±0.4}    &                      & \multicolumn{2}{c}{94.7}     & \multicolumn{3}{c}{±0.4}    &                      & \multicolumn{2}{c}{94.4}     & \multicolumn{3}{c}{±0.4}    &                      & \multicolumn{2}{r}{1653} & \multicolumn{2}{r}{\textless{}0.001} \\ \hline
\multicolumn{3}{l}{Vertical model}              & Overall          & \multicolumn{2}{l}{}        &                      & \multicolumn{5}{l}{User's}                                 &                      & \multicolumn{5}{l}{Producer's}                             &                      & \multicolumn{4}{l}{McNemar's test}                              \\ \cline{4-6} \cline{8-12} \cline{14-18} \cline{20-23} 
                     & \multicolumn{2}{l}{}     & accuracy         & \multicolumn{2}{r}{95\% CI} &                      & \multicolumn{2}{l}{accuracy} & \multicolumn{3}{r}{95\% CI} &                      & \multicolumn{2}{l}{accuracy} & \multicolumn{3}{r}{95\% CI} &                      & \multicolumn{2}{r}{Chi2} & \multicolumn{2}{l}{p-value}          \\ \hline
\multicolumn{4}{l}{Random   forest}                                & \multicolumn{2}{l}{}        &                      & \multicolumn{2}{l}{}         & \multicolumn{3}{l}{}        &                      & \multicolumn{2}{l}{}         & \multicolumn{3}{l}{}        &                      & \multicolumn{2}{l}{}     & \multicolumn{2}{l}{}                 \\
                     & High             & \multicolumn{2}{l}{}     & \multicolumn{2}{l}{}        &                      & \multicolumn{2}{c}{95.3}     & \multicolumn{3}{c}{±2.2}    &                      & \multicolumn{2}{c}{61.2}     & \multicolumn{3}{c}{±2.7}    &                      & \multicolumn{2}{r}{—}    & \multicolumn{2}{l}{}                 \\
                     & Low              & \multicolumn{2}{l}{}     & \multicolumn{2}{l}{}        &                      & \multicolumn{2}{c}{53.2}     & \multicolumn{3}{c}{±1.2}    &                      & \multicolumn{2}{c}{86.2}     & \multicolumn{3}{c}{±1.1}    &                      & \multicolumn{2}{r}{—}    & \multicolumn{2}{l}{}                 \\
                     & Not   built-up   & \multicolumn{2}{r}{80.1} & \multicolumn{2}{c}{±0.5}    &                      & \multicolumn{2}{c}{95}       & \multicolumn{3}{c}{±0.4}    &                      & \multicolumn{2}{c}{79.6}     & \multicolumn{3}{c}{±0.6}    &                      & \multicolumn{2}{r}{—}    & \multicolumn{2}{l}{}                 \\
\multicolumn{6}{l}{Fully convolutional network}                                                  & \multicolumn{1}{c}{} & \multicolumn{2}{c}{}         & \multicolumn{3}{c}{}        &                      & \multicolumn{2}{c}{}         & \multicolumn{3}{c}{}        &                      & \multicolumn{2}{r}{}     & \multicolumn{2}{r}{}                 \\
                     & High             & \multicolumn{2}{r}{}     & \multicolumn{2}{c}{}        & \multicolumn{1}{c}{} & \multicolumn{2}{c}{96.9}     & \multicolumn{3}{c}{±1.4}    &                      & \multicolumn{2}{c}{58.7}     & \multicolumn{3}{c}{±3.0}    &                      & \multicolumn{2}{r}{1}    & \multicolumn{2}{r}{0.4}              \\
                     & Low              & \multicolumn{2}{r}{}     & \multicolumn{2}{c}{}        & \multicolumn{1}{c}{} & \multicolumn{2}{c}{70.3}     & \multicolumn{3}{c}{±1.3}    &                      & \multicolumn{2}{c}{78.7}     & \multicolumn{3}{c}{±1.2}    &                      & \multicolumn{2}{r}{665}  & \multicolumn{2}{r}{\textless{}0.001} \\
                     & Not   built-up   & \multicolumn{2}{r}{87.8} & \multicolumn{2}{c}{±0.4}    & \multicolumn{1}{c}{} & \multicolumn{2}{c}{93.4}     & \multicolumn{3}{c}{±0.4}    &                      & \multicolumn{2}{c}{92.7}     & \multicolumn{3}{c}{±0.4}    &                      & \multicolumn{2}{r}{716}  & \multicolumn{2}{r}{\textless{}0.001} \\
\multicolumn{2}{l}{DeepLab}             & \multicolumn{2}{l}{}     & \multicolumn{2}{l}{}        &                      & \multicolumn{2}{l}{}         & \multicolumn{3}{l}{}        &                      & \multicolumn{2}{l}{}         & \multicolumn{3}{l}{}        &                      & \multicolumn{2}{l}{}     & \multicolumn{2}{l}{}                 \\
                     & High             & \multicolumn{2}{l}{}     & \multicolumn{2}{l}{}        &                      & \multicolumn{2}{c}{81.9}     & \multicolumn{3}{c}{±2.3}    &                      & \multicolumn{2}{c}{82.9}     & \multicolumn{3}{c}{±2.3}    &                      & \multicolumn{2}{r}{10}   & \multicolumn{2}{r}{0.001}            \\
                     & Low              & \multicolumn{2}{l}{}     & \multicolumn{2}{l}{}        &                      & \multicolumn{2}{c}{72.7}     & \multicolumn{3}{c}{±1.2}    &                      & \multicolumn{2}{c}{81.8}     & \multicolumn{3}{c}{±1.1}    &                      & \multicolumn{2}{r}{939}  & \multicolumn{2}{r}{\textless{}0.001} \\
                     & Not built-up     & \multicolumn{2}{r}{88.9} & \multicolumn{2}{c}{±0.4}    &                      & \multicolumn{2}{c}{95.1}     & \multicolumn{3}{c}{±0.4}    &                      & \multicolumn{2}{c}{91.4}     & \multicolumn{3}{c}{±0.5}    &                      & \multicolumn{2}{r}{868}  & \multicolumn{2}{r}{\textless{}0.001} \\ \hline
\end{tabular}}
\end{table*}

\subsection{Spatial and temporal transferability}
The DeepLab model had more robust results when using at least 30\% of the training datasets (Fig.~\ref{fig:samplesize}). Otherwise, the overall accuracy dropped rapidly. Moreover, DeepLab achieved higher accuracies than FCN only if the datasets used were larger than 50\% for the horizontal task and 70\% for the vertical classification. In contrast, the RF model resulted in consistent accuracies between large and small sample sizes. 

\begin{figure*}[!b]
    \centering
        \includegraphics[width=0.9\textwidth]{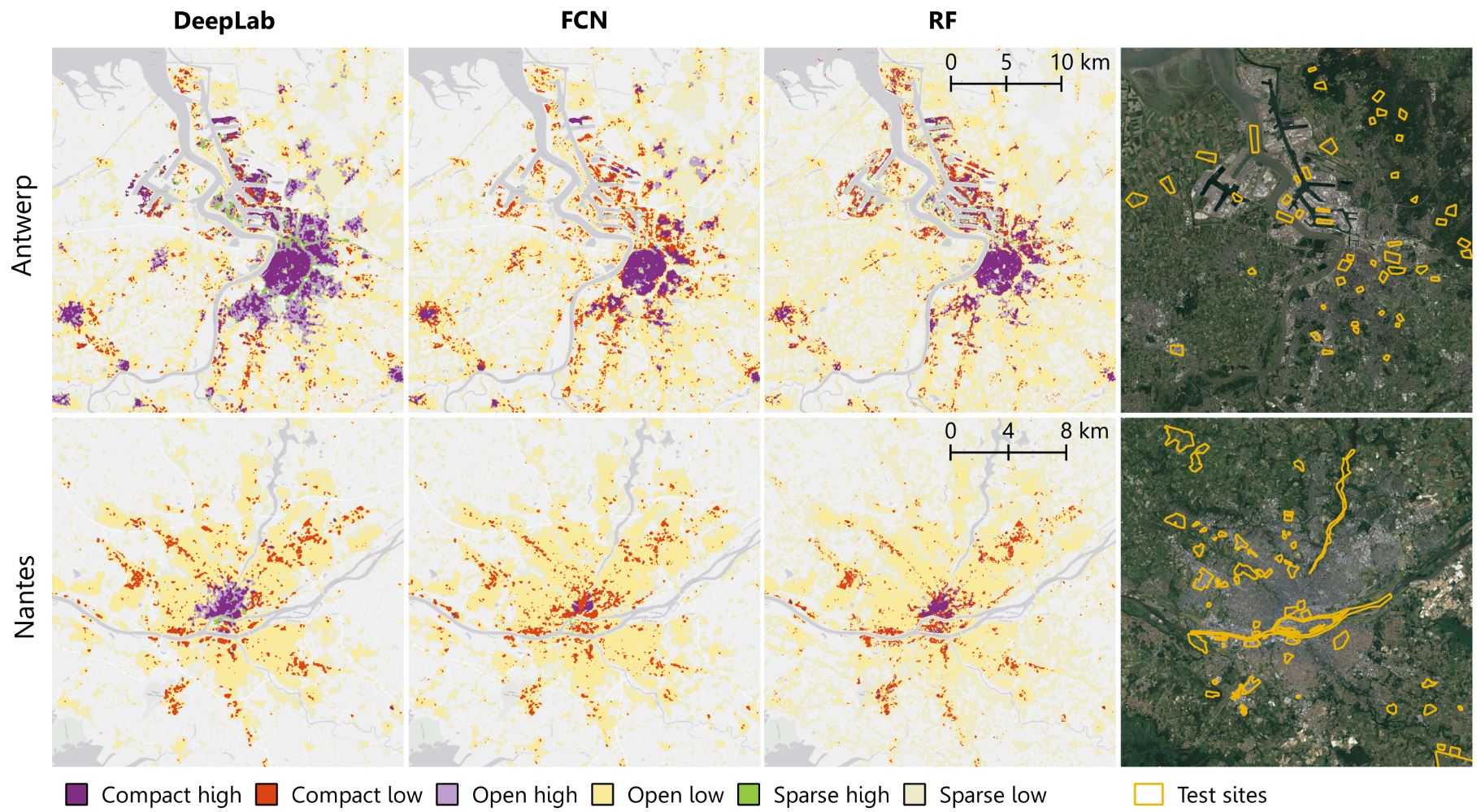}
    \caption{Comparison of DeepLab, FCN, and texture-based RF classifications for Antwerp and Nantes, mapped with Danish training data. Other cases are presented in Fig. S3-4, Appendix A.}
    \label{fig:EUcases}
\end{figure*}

When the models trained on Danish data were applied to the European cities outside Denmark, DeepLab outperformed in most of the cases (Table~\ref{tab:EUcities}). The overall accuracies of DeepLab ranged from 70\% (Arnhem) to 91.8\% (Antwerp), with an average value of 79.3\% in the horizontal dimension. For the vertical axis, the overall accuracies ranged from 69.3\% (Arnhem) to 89.6\% (Leuven), with an average of 81.4\%. The RF showed a similar difficulty as experienced for Denmark, namely that agricultural fields with grid patterns were mapped as \textit{sparse} settlements (e.g., the outskirts of Antwerp and Nantes, Fig.~\ref{fig:EUcases}). DeepLab had the highest producer’s accuracy for high-rise buildings (Table S6, Appendix A), while RF and FCN detected few \textit{high-rise} buildings with middle density (i.e., \textit{open}) in the suburbs (Fig.~\ref{fig:EUcases}).

\begin{table*}[!b]
\caption{Accuracies and 95\% confidence intervals for horizontal and vertical thematic classes based on the test data for 1995 and 2006.}
\centering
\setlength{\tabcolsep}{5pt}
\centering
\label{tab:EUcities}
\resizebox{0.65\textwidth}{!}{%
\begin{tabular}{llllllllllllll}
\hline
\multirow{3}{*}{City} & \multicolumn{6}{c}{Horizontal   dimension}                                       &           & \multicolumn{6}{c}{Vertical   dimension}                                               \\ \cline{2-7} \cline{9-14} 
                      & \multicolumn{2}{c}{DeepLab}   & \multicolumn{2}{c}{FCN} & \multicolumn{2}{c}{RF} &           & \multicolumn{2}{c}{DeepLab}   & \multicolumn{2}{c}{FCN}       & \multicolumn{2}{c}{RF} \\ \cline{2-7} \cline{9-14} 
                      & OA            & F1            & OA              & F1    & OA         & F1        &           & OA            & F1            & OA            & F1            & OA         & F1        \\ \hline
Wageningen            & \textbf{79.8} & \textbf{0.68} & 75.2            & 0.61  & 68.3       & 0.52      & \textbf{} & \textbf{84.8} & \textbf{0.76} & 84.1          & 0.76          & 72.2       & 0.51      \\
Arnhem                & \textbf{70}   & \textbf{0.62} & 69.9            & 0.6   & 67.5       & 0.62      & \textbf{} & \textbf{69.3} & \textbf{0.61} & 67.8          & 0.6           & 68.5       & 0.6       \\
Brussels              & \textbf{84}   & \textbf{0.76} & 80.9            & 0.72  & 82.3       & 0.76      &           & 84.4          & 0.77          & \textbf{87.8} & \textbf{0.81} & 81.8       & 0.75      \\
Antwerp               & \textbf{91.8} & \textbf{0.89} & 88.3            & 0.85  & 78.3       & 0.76      & \textbf{} & \textbf{89.3} & 0.86          & 88.9          & \textbf{0.87} & 81.4       & 0.79      \\
Leuven                & \textbf{82.8} & \textbf{0.74} & 79              & 0.68  & 75.3       & 0.65      &           & 89.6          & 0.83          & \textbf{91.7} & \textbf{0.87} & 84.4       & 0.76      \\
Hamburg               & 75.9          & \textbf{0.67} & \textbf{76.5}   & 0.56  & 61.8       & 0.51      & \textbf{} & \textbf{83.6} & \textbf{0.79} & 79            & 0.73          & 75.7       & 0.7       \\
Berlin                & \textbf{77.8} & \textbf{0.68} & 76.6            & 0.67  & 73.3       & 0.64      & \textbf{} & \textbf{85.8} & \textbf{0.82} & 79.4          & 0.72          & 78.7       & 0.72      \\
Nantes                & \textbf{72.3} & \textbf{0.69} & 67.3            & 0.63  & 65.5       & 0.64      & \textbf{} & \textbf{71.9} & \textbf{0.69} & 63.4          & 0.58          & 65.1       & 0.64      \\
Augsburg              & \textbf{87}   & \textbf{0.82} & 84.6            & 0.77  & 81.1       & 0.78      & \textbf{} & \textbf{82.4} & \textbf{0.76} & 80.6          & 0.72          & 76.6       & 0.66      \\
Paris                 & \textbf{71.2} & \textbf{0.69} & 70.2            & 0.67  & 69         & 0.65      & \textbf{} & \textbf{73.2} & \textbf{0.72} & 59.2          & 0.48          & 67.9       & 0.65      \\ \hline
\end{tabular}}
\end{table*}

To evaluate the temporal generalization of DeepLab, FCN, and RF for time series analysis, we trained a model with the data from 2014 and applied it to classify urban density for 1995 and 2006. Semantic segmentation models (DeepLab and FCN) had higher accuracies in the previous test years, compared to texture-based RF (Table~\ref{tab:annual}). All three models reached high accuracy (approximately 90\%) for the training year 2014 as expected. However, the accuracy of the RF classifications in 2006 dropped by 12 percentage points for the horizontal and 7 percentage points for the vertical direction. The performance of the RF decreased even more once we classified 1995 urban form with models trained on the 2014 data, by 15 and 10 percentage points in the horizontal and vertical directions. In contrast, DeepLab and FCN models achieved more consistent accuracies between 1995, 2006, and 2014, with a small difference of around 1 percentage point. 

\vspace{0.1cm}
\begin{table}[H]
\caption{Accuracy comparison across years (unit: \%).} 
\centering
\setlength{\tabcolsep}{5pt}
\centering
\label{tab:annual}
\resizebox{\columnwidth}{!}{%
\begin{tabular}{llllllll}
\hline
        & \multicolumn{3}{c}{OA in horizontal dimension} &           & \multicolumn{3}{c}{OA in vertical dimension}  \\ \cline{2-4} \cline{6-8} 
        & 1995           & 2006          & 2014          &           & 1995          & 2006          & 2014          \\ \hline
DeepLab & \textbf{89.5}  & \textbf{88.7} & \textbf{89.1} & \textbf{} & \textbf{91.8} & \textbf{91.4} & \textbf{92.1} \\
FCN     & 88.6           & 87.6          & 89            &           & 89.5          & 90.2          & 90.9          \\
RF      & 72             & 75.6          & 87.4          &           & 78.4          & 81.8          & 88.7          \\ \hline
\end{tabular}}
\end{table}

We assessed the accuracy of urban form change in pixels experiencing densification between 2006 and 2014 (Table~\ref{tab:growth}). Overall, the models more accurately detected density growth in the horizontal dimension than the vertical dimension. For horizontal densification, the three models reached similar producer accuracies of 88-89\% in highly built-up regions. For less dense areas, however, DeepLab and FCN were more accurate than the RF model to detect horizontal densification. DeepLab also best depicted vertical growth (Table~\ref{tab:growth}).

\vspace{0.1cm}
\begin{table*}[!b]
\caption{Accuracy comparison of density growth between 2006 and 2014 detected by DeepLab, FCN, and RF.}
\centering
\setlength{\tabcolsep}{5pt}
\centering
\label{tab:growth}
\resizebox{0.75\textwidth}{!}{%
\begin{tabular}{llllllllllll}
\hline
        & \multicolumn{3}{l}{Growth in vertical density} &  & \multicolumn{3}{l}{\begin{tabular}[c]{@{}l@{}}Growth in horizontal density \\    (low density developed area)\end{tabular}} &  & \multicolumn{3}{l}{\begin{tabular}[c]{@{}l@{}}Growth in horizontal density \\    (high density developed area)\end{tabular}} \\ \cline{2-4} \cline{6-8} \cline{10-12} 
        & UA(\%)         & PA(\%)         & F1           &  & UA(\%)                                   & PA(\%)                                   & F1                                    &  & UA(\%)                                   & PA(\%)                                   & F1                                     \\ \hline
DeepLab & 59.9           & 70.8           & 0.65         &  & 54.0                                     & 76.4                                     & 0.63                                  &  & 66.6                                     & 88.9                                     & 0.76                                   \\
FCN     & 54.9           & 69.5           & 0.61         &  & 52.4                                     & 76.6                                     & 0.62                                  &  & 60.9                                     & 89.5                                     & 0.72                                   \\
RF      & 45.1           & 58.6           & 0.51         &  & 35.9                                     & 50.1                                     & 0.42                                  &  & 62.2                                     & 88.3                                     & 0.73                                   \\ \hline
\end{tabular}}
\end{table*}

\begin{figure*}[!b]
    \centering
        \includegraphics[width=0.9\textwidth]{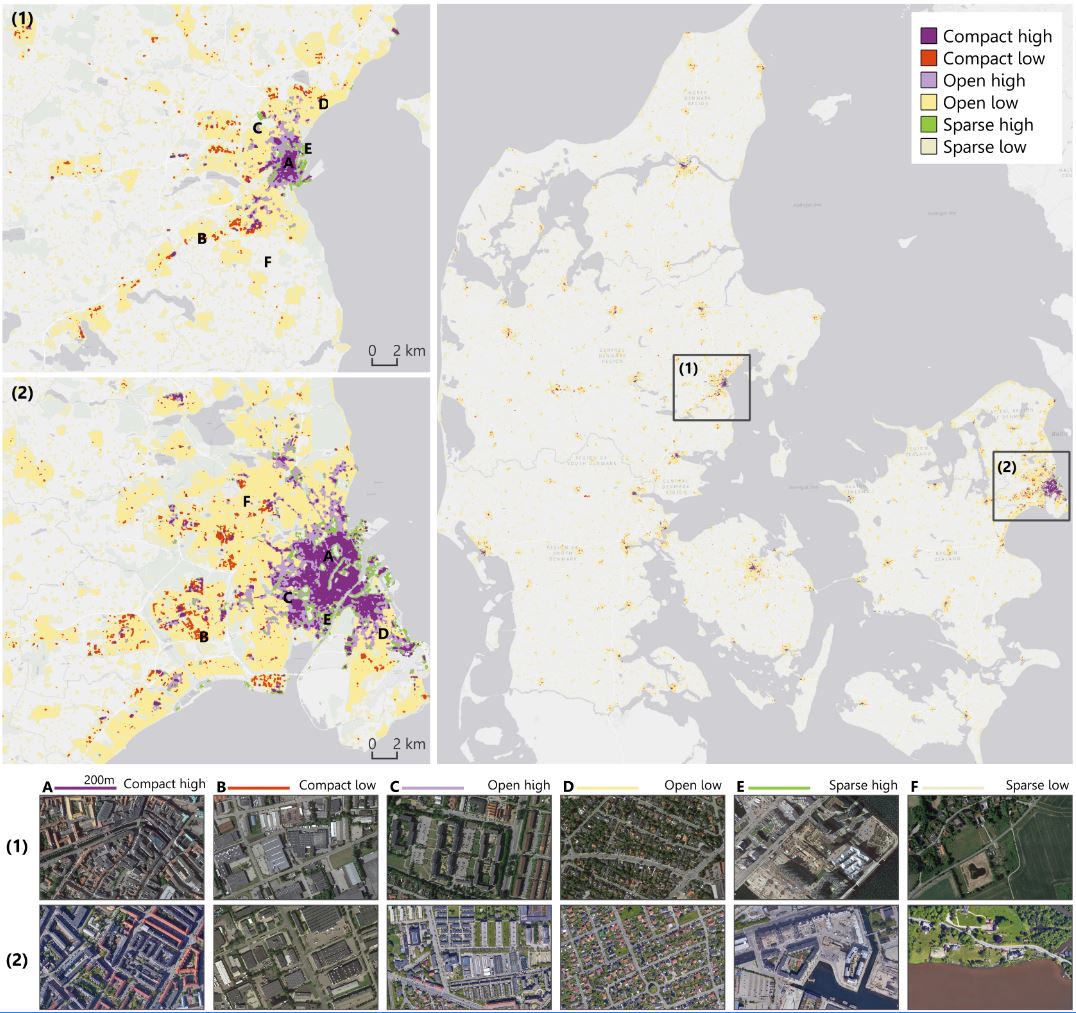}
    \caption{The results of the 2014 urban density map in (1) Aarhus and (2) Copenhagen in two morphological dimensions:  horizontal (compact, open, sparse) and vertical (high, low). The detailed sites of each class, labeled by A-F, show the coherence of different types of the neighborhood in comparison with Google Earth images.}
    \label{fig:map2014}
\end{figure*}

\subsection{Contemporary urban form patterns at regional and national scales}

Fig.~\ref{fig:map2014} presents the final urban form map combining the two dimensions with insets of two large cities, Aarhus and Copenhagen. Visual inspection of classifications reveals a coherent class assignment without artifacts, such as isolated misclassified pixels. The detected spatial patterns of the urban form appear meaningful and follow typical patterns of the Danish urban development along the radial railway corridors. The CBDs are primarily classified as \textit{compact-high} buildings, followed with \textit{open-high, open-low, sparse-low} arrangements along the urban-rural corridors. In Copenhagen and Aarhus, \textit{compact-low} neighborhoods are mainly located in the satellite suburbs instead of the central district. Sparse-high rises predominantly occur close to the harbors, characterized by modern architecture. The \textit{compact-high} zones mainly consist of apartment buildings arranged in rows or large blocks, while the \textit{open-high} zones represent residential row apartments surrounded by green space. Large residential areas are classified as \textit{open-low}, where duplex or single-family houses occur with home gardens. Other \textit{compact-low} zones, which have been outsourced to the suburbs, represent industrial areas that were either in use or remodeled for residential or recreational purposes. In sum, the spatial patterns of urban density classification show the qualitative logic of spatial planning and urban life.

\begin{figure*}[!b]
    \centering
        \includegraphics[width=0.8\textwidth]{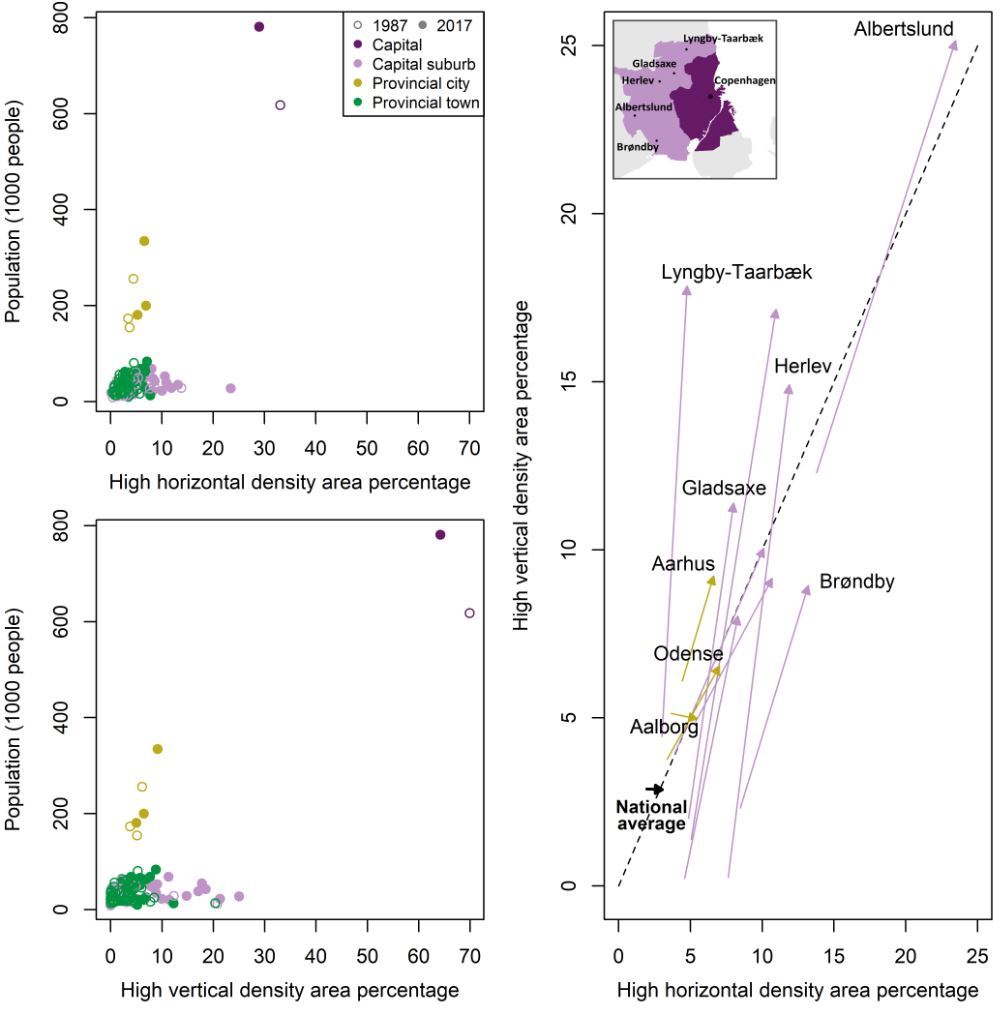}
    \caption{Changes of population and the percentage of high building density areas among the total built-up areas in horizontal and vertical dimensions, in the capital city, capital suburbs, provincial cities, and provincial towns between 1987 and 2017. Arrows represent changes in urban density.}
    \label{fig:popXform}
\end{figure*}

\subsection{Trends of building growth versus population growth }

\begin{figure*}[!t]
    \centering
        \includegraphics[width=0.9\textwidth]{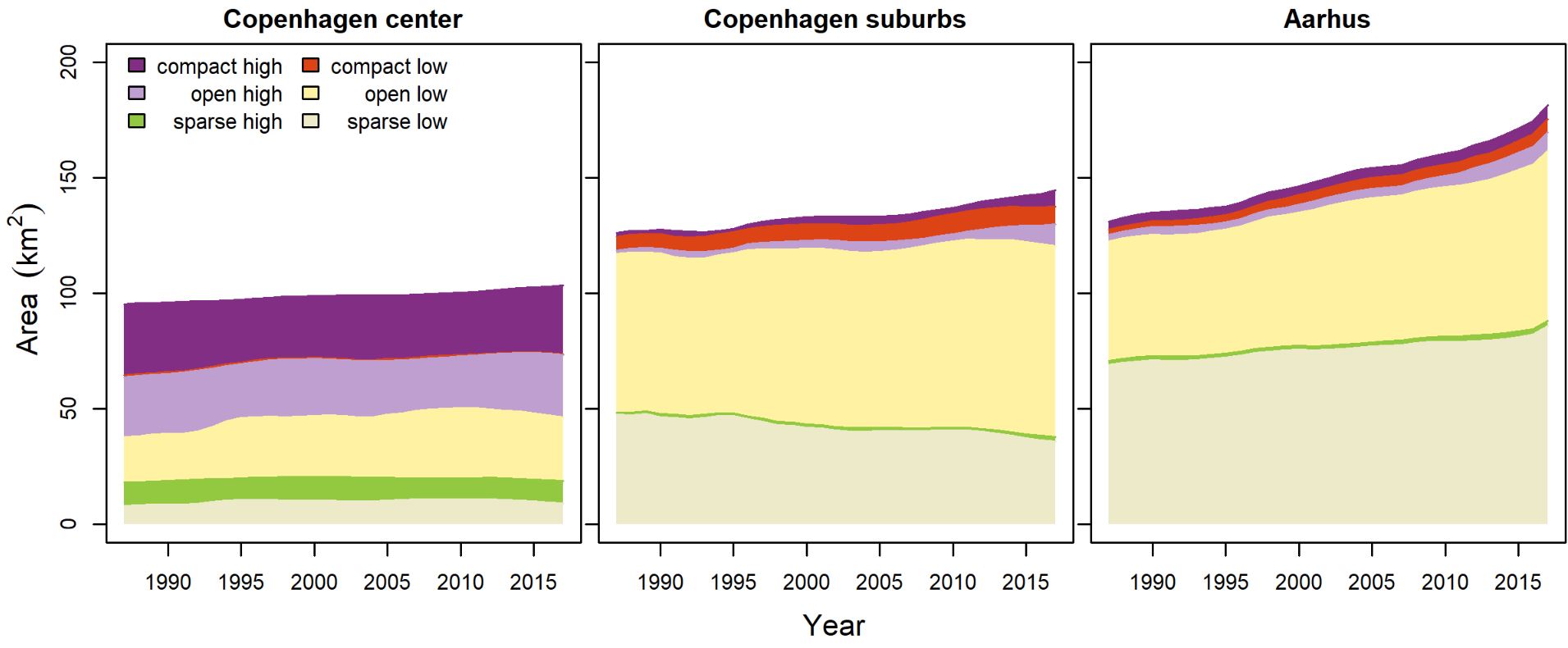}
    \caption{Urban form dynamic from 1987 to 2017 in the capital, Copenhagen, its suburbs, and the second highest populated city Aarhus.}
    \label{fig:trend}
\end{figure*}

In 1987, Denmark had approximately 13,624 ha of \textit{high-rise} areas and 9,247 ha of \textit{compact} neighborhoods, from a total 472,328 ha of built-up areas. By 2017, the country’s \textit{high-rise} area had grown to approximately 21,160 ha, and the \textit{compact} area had increased to 20,646 ha. The average annual rate of dense housing growth in the horizontal dimension (380 ha/year) remains higher than the growth of \textit{high-rise} buildings (251 ha/year). A rising population is commonly recognized as one of the main drivers of urban densification. In Denmark, however, densification does not occur in the cities with the most population growth. Various temporal dynamics and different forms of building constructions are revealed between the capital, capital suburbs, and provincial cities. 

Over the 31 years, the capital and other primary cities have high population growth but low dense-housing growth (Fig.~\ref{fig:popXform}). The most significant discrepancy is in central Copenhagen. From 1987 to 2017, the population of the capital increased by 26\% (from 618,038 to 781,089). However, most of the housing was developed decades ago, and little new development has been built since 1987 (Fig.~\ref{trend}), indicating a shrinkage of living space for the downtown residents. In the second most populated city Aarhus, the dense built-up also has not grown hand-in-hand with population growth. Neighborhoods with spacious open space increase among all types of urban form (Fig.~\ref{trend}). In contrast, the vertical urban growth and horizontal density growth in Denmark mainly occur in the capital suburbs, where the population remains less than 100,000. In those satellite suburbs such as Herlev, Gladsaxe, and Albertslund, which have been developed to tackle the population pressure of central Copenhagen, the vertical growth exceeds the horizontal densification (Fig.~\ref{fig:popXform}).

\section{Discussion}
\subsection{Monitoring emergence and transformation of horizontal and vertical urban form}

The remote sensing and urban planning communities are calling for consistent urban form classification, in both horizontal and vertical directions (\cite{Wentz2018}). \cite{Frolking2013} made the first attempts to measure changes in the urban structure of 100 large cities on different continents. They compared urban growth in both two and three dimensions at around 5km resolution by using SeaWinds microwave and nighttime light data. Further progress was made by \cite{Mahtta2019}, who up-scaled the comparison to all global cities with a population of more than 1 million (i.e., 499 cities) at a similar mapping resolution. Although the majority of models have been developed in large cities, the most of future urban growth will occur in small- and medium-sized cities and towns with population of fewer than 1 million (\cite{Reba2020}). The mapping spatial resolution challenges the monitoring of small-scale urban growth. At resolution lower than 1km, the detailed patterns within a medium-sized city, such as urban densification, infill development, urban renewal, are aggregated and invisible. 

Here, we demonstrate that a semantic segmentation workflow, unlike typical scene classification, can prevent loss of spatial resolution of the Landsat time series. The accuracies achieved by DeepLab and FCN are significantly higher and more balanced between classes compared to a texture-based RF that has been commonly used in recent LCZ studies (\cite{Bechtel2016,Wieland2016,Xu2017}). Our results show that DeepLab, a semantic segmentation model enhanced by an atrous spatial pyramid pooling structure, is more accurate than a simple semantic segmentation method FCN. Compared to FCN considering spatial-spectral information within one and two neighboring pixels (before and after pooling) away from the center, DeepLab’s atrous pooling takes multiple neighboring profiles from 2, 4, 8 pixels (atrous rates 1, 2, 4 when downsampling size to H/2, W/2) plus from 48 pixels for a patch-level feature. The influence of patch size and atrous rates on the prediction performance might be explained by what we need for urban form prediction, specifically from the Landsat images. For instance, the local built-environment characteristics within 8 pixels (240 m) and a regional urban-rural context captured by a 48 $\times$ 48 patch (around 1.5 km) can be useful to learn the high rise class with different backgrounds. In comparison, a 20 $\times$ 20 patch might be insufficient to predict such information. We note that the advantage of DeepLab over FCN may only appear when utilizing a large training dataset. 

Our results show convincing accuracies for a wide diversity of urban form at 30m spatial resolution in Denmark, where small-scale urban densification has been occurring. The classifications from 1987 to 2017 reveal distinct urban growth patterns. The clusters of four types of cities (i.e., the capital, capital suburbs, provincial cities, and provincial towns) present varying rationales of urban form development. For instance, although the capital and other major Danish cities experienced post World War II suburbanization, high-rise buildings have increased in capital suburbs at a higher pace and extent compared to in other cities. At the same time, the capital’s downtown has a shortage of land but a de-densification process in housing, particularly along its waterfront, such as at Sydhavn. While the housing density has decreased, the population continues to increase in the capital, indicating that living space per capita is shrinking. The result in Copenhagen runs counter to a global-scale finding that urban growth is becoming more expansive rather than compact (\cite{Seto2011}). While the transformation of horizontal and vertical density is linked to an increase in housing prices, the dynamic could also influence the living quality and well-being that the neighborhoods promote (\cite{Haaland2015,Melis2015}). Thus, it is crucial to evaluate the effects of urban dynamics to support urban development policies and provide better solutions for the future. 

\subsection{Spatial and temporal transferability for long-term mapping}

Urban form remote sensing studies often emphasize the transferability of approaches. Many have contributed to spatial generalization, indicating where a model trained on data from one region can be successfully applied to another using Sentinel-2 data (\cite{Qiu2020fcn,Rosentreter2020}). For Landsat-based LCZ mapping, \cite{Demuzere2019} found that the transferability was rather tricky. The authors show that LCZ’s overall accuracies (OA) of the RF model were high ($>$80\%) when using training and test data in the same city, but accuracies for city-to-city transferability tests were lower than 40\%. Even for transferability tests within the same ecoregion, such as temperate forest Europe (tests between Paris, London, Warsaw), OA still varied between 40\% and 70\%. Yoo et al. (2019) showed that the CNN model could improve the transferability of Landsat-based LCZ mapping (OA ranged from 42\% to 78\%) compared to the RF classifier (27\% to 61\%). Our study presents that the DeepLab model, using atrous spatial pyramids, is more transferable (OA settled between 70\% and 92\% for the horizontal and 69\% to 90\% for the vertical) to other European cities outside Denmark, compared to the simple CNN model (averagely 3\% lower) and texture-based RF (6\% lower), indicating that DeepLab can ameliorate transferability of three-dimensional urban form mapping. 

We demonstrate the utility of temporal transferability — where models trained with annotations from a certain period can be used to map another past period. This transferability is vital when historical reference data is not available or difficult to collect. The results show that our deep learning-based models could not only accurately map the initial training year (i.e., 2014), but also generalize to map urban form back in time (i.e., 2006 and 1995). In contrast, the texture-based RF classifier is limited in temporal transferability when the time of prediction is far from the training year. Therefore, we propose that the semantic segmentation models could be applied to map urban dynamics retrospectively more successfully compared to the non-deep machine learning model. One of the reasons may be that CNN-based models are using deep hidden layers to capture spatial context, which is more representative than the user-defined texture features used in RF classifiers, such as the first- and second-order statistics (\cite{Rosentreter2020}). Another explanation is that DeepLab approaches using spatial pyramid pooling are able to utilize spatial profiles from a small to a larger scale (\cite{Chen2017}). Those multi-scale features could be useful to mimic the spatial configuration of cities at different times when the urban development scale has shrunk or expanded.

Multi-temporal records of urban form are crucial for scientists and planners seeking evidence-based inference and solutions. Epidemiologists have found that the characteristic of the residential area in people’s early life could explain their health outcomes later in life (\cite{Curtis2004}). Another study of dengue outbreaks also observed synergistic effects between housing conditions and precipitation over more than a decade (\cite{Chen2018dengue}). For other environmental goals, such as air quality, scientists also endeavor to evaluate urban morphology effects with longitudinal analysis (\cite{Bereitschaft2013}). Although those studies measured environmental factors, including air, rainfall, temperature, with multiple time steps, the 3D urban form databases were uni-temporal. Such inferential approaches were based on the assumption that urban density was static. Nevertheless, even in Denmark, a developed Nordic country, the densification appears at a variety of speeds and dimensions. It is an opportunity for the remote sensing community to engage in the strategies of sustainable development by providing urban dynamics products and to relate earth observations to other aspects of human well-being.  

\subsection{Limitations and outlook}

In this study, we explored the suitability of CNN-based semantic segmentation models for the retrospective reconstruction of urban dynamics. For the selected approaches, the requirement to prepare extensive training data may pose some limitations. For future studies, achieving high performance with limited training data using our method will be cost-effective for multi-temporal urban mapping. Nevertheless, a challenge remains to use time-series data analysis based on satellite imagery to make consistent comparisons over a study period. Harmonization of images from Landsat TM, ETM+, and OLI has been developed for a comparable vegetation index (\cite{Roy2016,Vogeler2018}). It is vital to mature such calibration for monitoring urban growth, which is more reflective in short-wave infrared and visible bands. Regarding the model, instead of using CNN-based models alone, hybrid models stacking convolutional and recurrent networks are emerging for deep-learning-based time-series image analysis (\cite{Mou2018,Qiu2020,Russwurm2018}). It would be interesting to explore whether those models could further improve urban dynamics reconstruction. Temporally dense reference data could also provide a more in-depth investigation regarding temporal transferability and temporal accuracy, such as evaluating annual change rates. Finally, although the urban density is defined by physical and geometrical characteristics, which are supposed to be impartial to cultural, environmental, and governance contexts, the performance of the models might differ between societies. Thus, further evaluation of the effectiveness of using deep learning methods with Landsat data beyond a national scale is needed. Despite some limitations, this study adds a baseline for an urban application of the Landsat program in the era of artificial intelligence.  

\section{Conclusions}
This study investigated the effectiveness of CNN-based semantic segmentation approaches incorporating Landsat satellite data to provide multi-annual urban form maps. Our findings confirm that the semantic segmentation models could significantly improve accuracies using Landsat imagery for providing 30m resolution urban form maps in both horizontal and vertical dimensions. DeepLab and FCN outperform a texture-based RF model in terms of the spatial and temporal transferability and could benefit analysis over a long period, and DeepLab better depicts vertical urban growth. We demonstrate the usefulness of long-term urban form maps that allow characterizing different pathways of urban growth in a variety of cities in Denmark. Notably, during the period 1987 to 2017, the urban density in the capital’s center has slightly decreased along with small-scale renewal projects. On the other hand, high-rise buildings have dominated the newly developed areas in the capital’s suburbs. The densification trend in the capital’s suburbs is even faster than in the second-largest city, Aarhus. Our proposed workflow based on Landsat time series and the CNN-based semantic segmentation can likely be replicated elsewhere to reconstruct urban dynamics.

\section*{Acknowledgement}
This study was supported by a Ph.D. scholarship from the Ministry of Education, Taiwan, by BERTHA - the Danish Big Data Centre for Environment and Health funded by the Novo Nordisk Foundation Challenge Programme (grant NNF17OC0027864), and by the European Research Council (ERC) under the European Union’s Horizon 2020 research and innovation program (Grant Agreement No. ERC-2016-StG-714087, Acronym: So2Sat). The work of Chunping Qiu was supported by the China Scholarship Council (CSC). The authors wish to thank the editor and four anonymous reviewers for constructive reviews and Yu-Te Chiang for supporting graphic design.

\section*{CRediT author statement}
\textbf{Tzu-Hsin Karen Chen}: Conceptualization, Methodology, Data Curation, Software, Formal analysis, Validation, Visualization, Writing - original draft. \textbf{Chunping Qiu}: Methodology, Data Curation, Software, Validation, Writing - review \& editing. \textbf{Michael Schmitt}: Methodology, Supervision, Writing - review \& editing. \textbf{Xiao Xiang Zhu}: Supervision, Resources, Writing - review \& editing. \textbf{Clive E. Sabel}: Supervision, Resources, Funding acquisition, Writing - review \& editing. \textbf{Alexander V. Prishchepov}: Conceptualization, Resources, Supervision, Writing - review \& editing.

\section*{References}
\printbibliography[title={references},heading=none]

\clearpage



\section*{Supplementary material}

Mapping horizontal and vertical urban densification in Denmark with Landsat time-series from 1985 to 2018: a semantic segmentation solution

Tzu-Hsin Karen Chen, Chunping Qiu, Michael Schmitt, Xiao Xiang Zhu, Clive E. Sabel, Alexander V. Prishchepov

**Supplementary data**

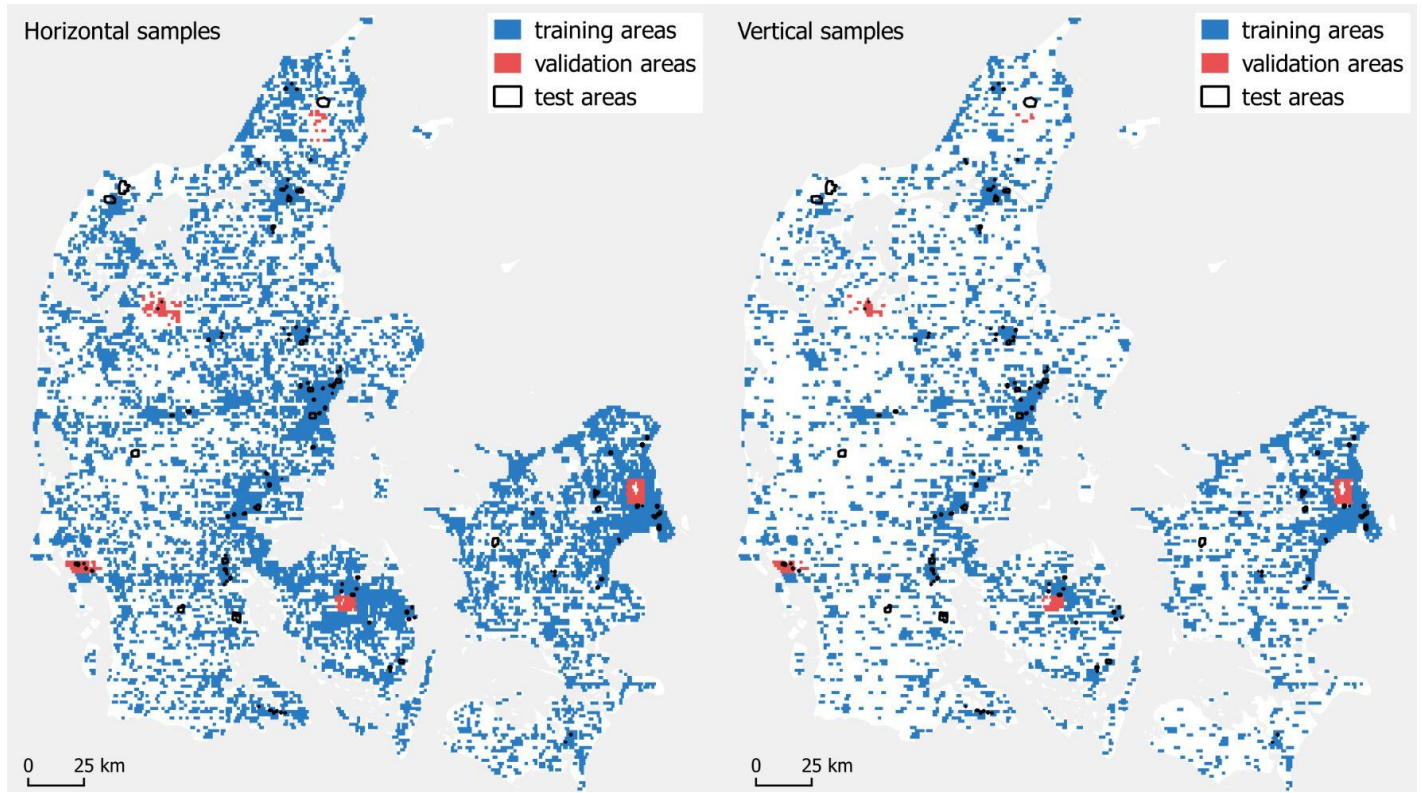

Fig. S1. The geography of training, validation, and test areas. Note that the test areas spatially overlapped with the training areas. The purpose was to compare the test accuracy from 2014 (when training data was used at the test sites) with the test accuracy from 1995 and 2006 (when images of the test sites were not used for training).

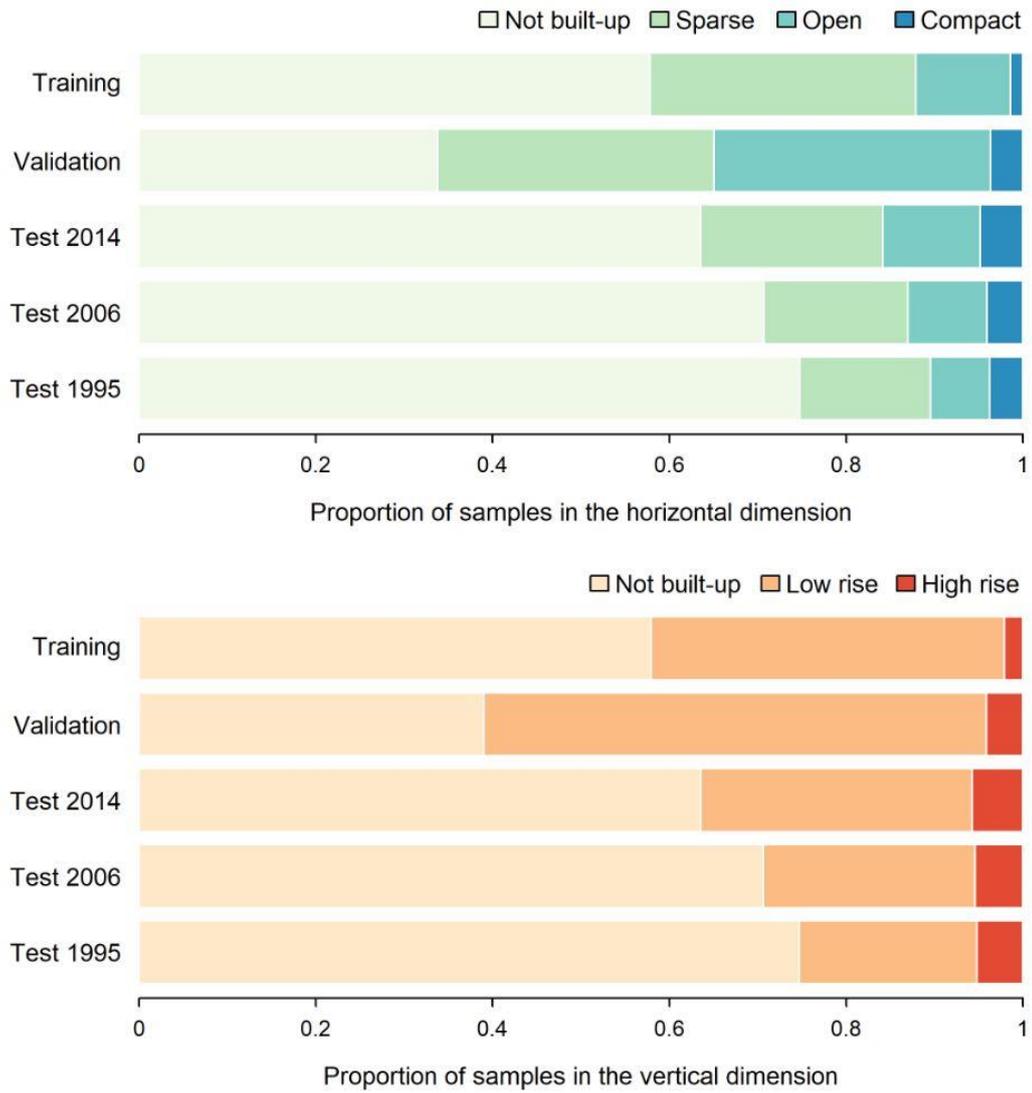

Fig. S2. Proportion of training, validation, and test data for the two classifications.

Table S1.

The corresponding LCZ labels for horizontal and vertical urban form classes, which were used for testing cities outside Denmark and comparing the spatial transferability of DeepLab, fully convolutional network, and random forest.

| ID | LCZ class | Horizontal class | Vertical class |
|---|---|---|---|
| 1 | Compact high | Compact | High |
| 2 | Compact mid-rise | Compact | High |
| 3 | Compact low-rise | Compact | Low |
| 4 | Open high-rise | Open | High |
| 5 | Open mid-rise | Open | High |
| 6 | Open low-rise | Open | Low |
| 7 | Lightweight low-rise | Compact | Low |
| 8 | Large low-rise | Compact | Low |
| 9 | Sparsely built | Sparse | Low |
| 10 | Heavy industry | Excluded | Excluded |
| A | Dense trees | Not built-up | Not built-up |
| B | Scattered trees | Not built-up | Not built-up |
| C | Bush, scrub | Not built-up | Not built-up |
| D | Low plants | Not built-up | Not built-up |
| E | Bare rock or paved | Not built-up | Not built-up |
| F | Bare soil or sand | Not built-up | Not built-up |
| G | Water | Excluded | Excluded |

Table S2.

Number of sample for testing European cities outside Denmark.

| City | Horizontal dimension | | | | | Vertical dimension | | | |
|---|---|---|---|---|---|---|---|---|---|
| | Compact | Open | Sparse | Not built-up | Total | High | Low | Not-built-up | Total |
| Wageningen | 115 | 200 | 73 | 487 | 875 | 112 | 300 | 487 | 899 |
| Arnhem | 200 | 200 | 0 | 600 | 1000 | 300 | 300 | 600 | 1200 |
| Brussels | 200 | 99 | 200 | 600 | 1099 | 119 | 300 | 600 | 1019 |
| Antwerp | 200 | 200 | 200 | 600 | 1200 | 300 | 300 | 600 | 1200 |
| Leuven | 200 | 181 | 96 | 600 | 1077 | 77 | 300 | 600 | 977 |
| Hamburg | 200 | 200 | 0 | 600 | 1000 | 300 | 300 | 600 | 1200 |
| Berlin | 200 | 200 | 200 | 600 | 1200 | 300 | 300 | 600 | 1200 |
| Nantes | 200 | 200 | 200 | 600 | 1200 | 300 | 300 | 600 | 1200 |
| Augsburg | 200 | 200 | 0 | 600 | 1000 | 300 | 300 | 600 | 1200 |
| Paris | 200 | 200 | 200 | 600 | 1200 | 300 | 300 | 600 | 1200 |

Table S3.

Results of random forest classifications with a range of window size for extracting texture features.

| Window size | Horizontal density model | | | | Vertical density model | | | |
|---|---|---|---|---|---|---|---|---|
| | OA | Kappa | Average F1 | Computation time (minute) | OA | Kappa | Average F1 | Computation time (minute) |
| $3 \times 3$ | 72.7% | 0.48 | 0.66 | 14 | 78.5% | 0.56 | 0.74 | 12 |
| **$5 \times 5$** | **73.3%** | **0.50** | **0.67** | 14 | **79.3%** | **0.57** | **0.75** | 12 |
| $7 \times 7$ | 67.0% | 0.41 | 0.62 | 14 | 74.4% | 0.50 | 0.73 | 13 |
| $47 \times 47$ | 59.0% | 0.26 | 0.52 | 16 | 67.9% | 0.36 | 0.69 | 14 |

Table S4. Results of semantic segmentation with varying patch sizes. OA: overall accuracy.

| Patch size | DeepLab for horizontal density model | | | | DeepLab for vertical density model | | | |
|---|---|---|---|---|---|---|---|---|
| | OA | Kappa | Average F1 | Computation time (minute) | OA | Kappa | Average F1 | Computation time (minute) |
| $20 \times 20$ | 87.3% | 0.72 | 0.76 | 509 | 88.6% | 0.74 | 0.82 | 294 |
| $40 \times 40$ | 87.1% | 0.71 | 0.76 | 121 | 88.2% | 0.73 | 0.84 | 56 |
| **$48 \times 48$** | 87.7% | 0.72 | 0.78 | 84 | 88.9% | 0.74 | 0.84 | 44 |
| $64 \times 64$ | 85.4% | 0.68 | 0.73 | 59 | 86.5% | 0.68 | 0.80 | 26 |
| Patch size | FCN for horizontal density model | | | | FCN for vertical density | | | |
| | OA | Kappa | Average F1 | Computation time (minute) | OA | Kappa | Average F1 | Computation time (minute) |
| $20 \times 20$ | 86.1% | 0.69 | 0.75 | 77 | 87.5% | 0.72 | 0.82 | 45 |
| $40 \times 40$ | 85.9% | 0.68 | 0.72 | 16 | 87.8% | 0.71 | 0.81 | 8 |
| $48 \times 48$ | 86.4% | 0.69 | 0.76 | 13 | 87.8% | 0.71 | 0.80 | 7 |
| $64 \times 64$ | 86.6% | 0.68 | 0.73 | 6 | 87.8% | 0.70 | 0.80 | 3 |

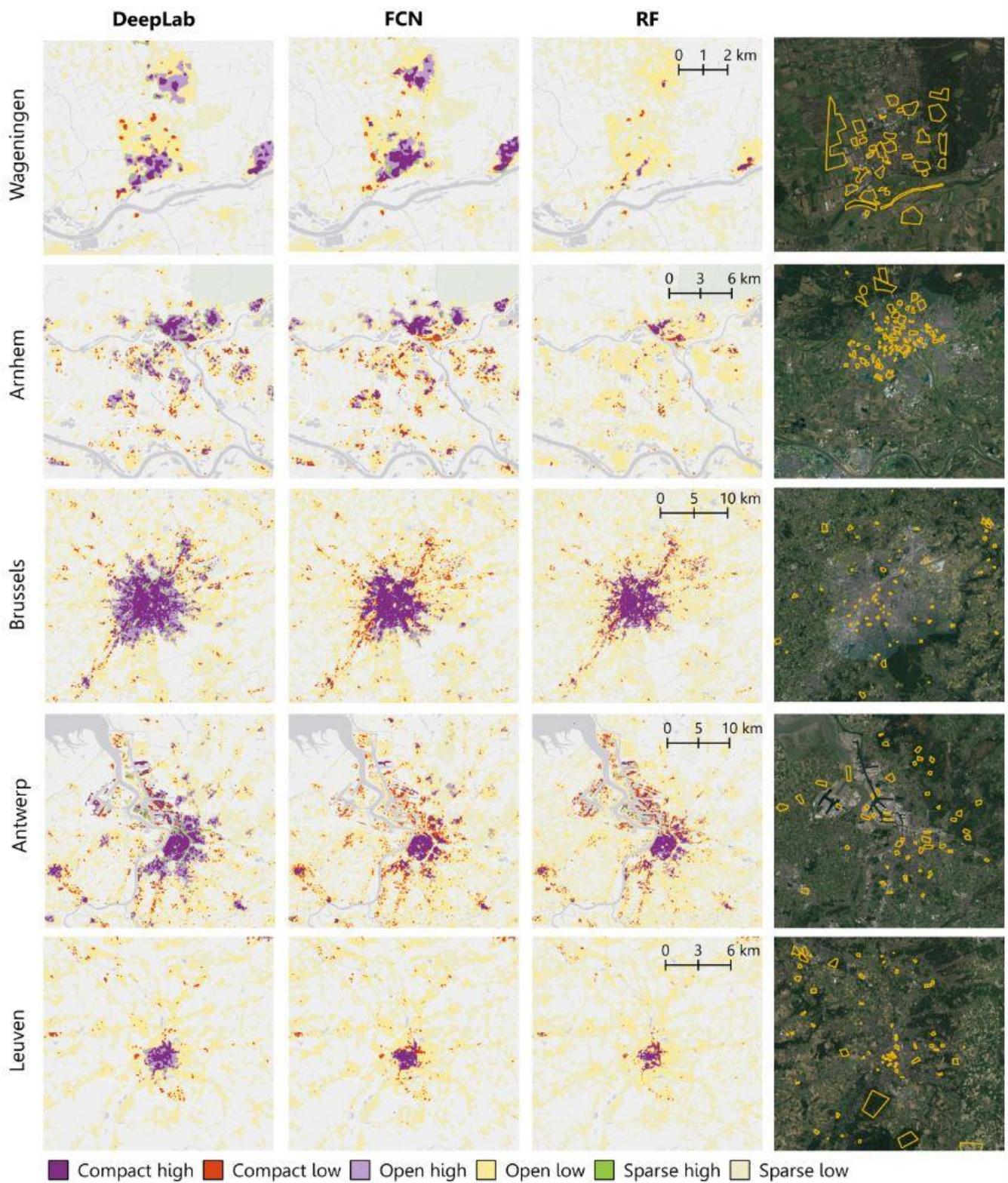

Fig. S3. Comparison of DeepLab, FCN, and texture-based RF classifications for spatial transferability tests (part 1/2).

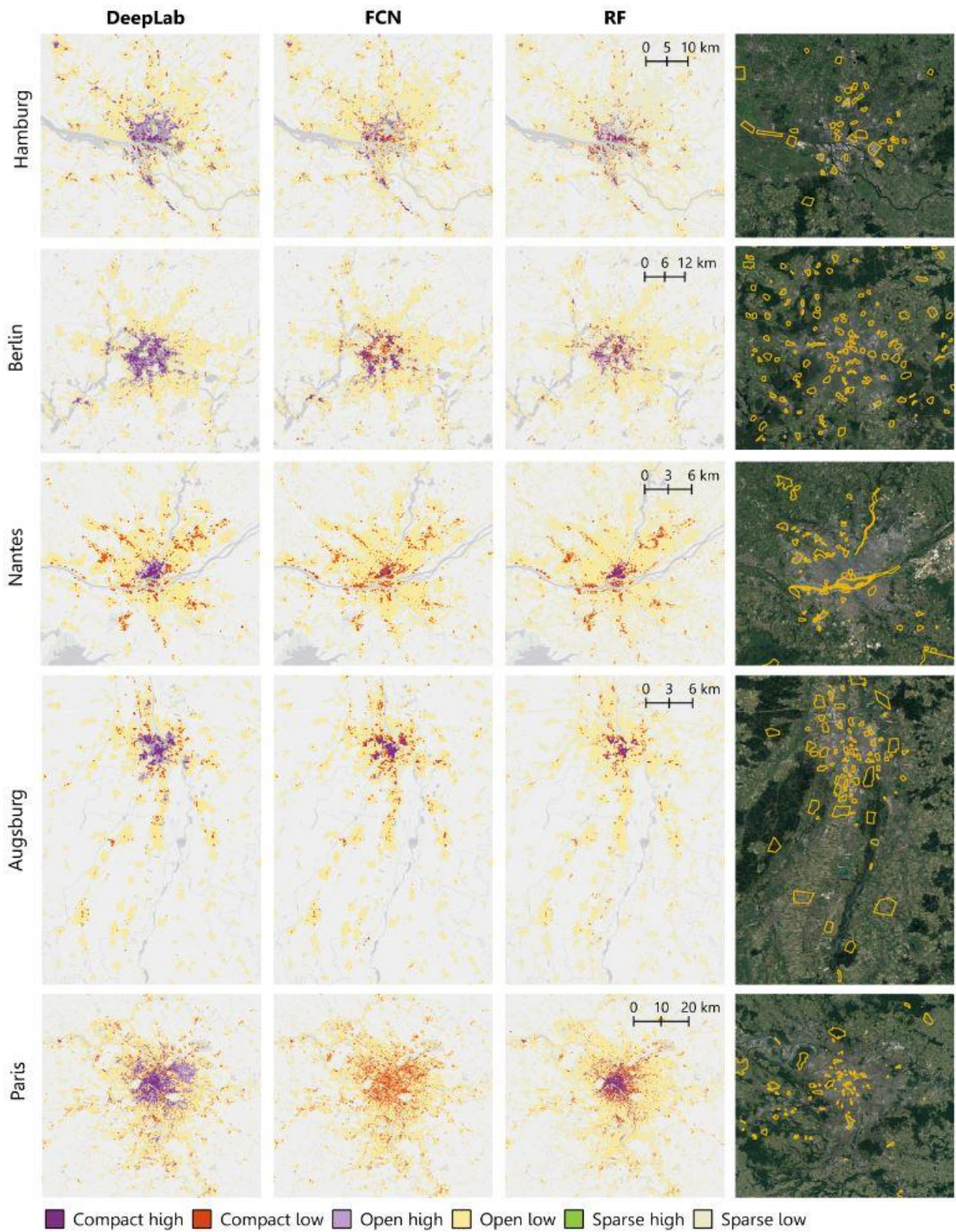

Fig. S4. Comparison of DeepLab, FCN, and texture-based RF classifications for spatial transferability tests (part 2/2).

Table S5. Spatial transferability tests on the trained models for horizontal density classification: Training samples from Danish cities used to train the models for mapping cities outside Denmark.

| Model | City | OA | Average F1 | Not built-up | | Sparse | | Open | | Compact | |
|---|---|---|---|---|---|---|---|---|---|---|---|
| | | | | UA | PA | UA | PA | UA | PA | UA | PA |
| DeepLab | Wageningen | 79.8 | 0.68 | 97.9 | 95.7 | 41.1 | 89.0 | 67.3 | 53.5 | 73.2 | 52.2 |
| | Arnhem | 70.0 | 0.62 | 90.7 | 87.8 | — | — | 52.5 | 42.5 | 59.5 | 44.0 |
| | Brussels | 84.0 | 0.76 | 100 | 94.0 | 71.1 | 70.0 | 41.9 | 78.8 | 92.8 | 70.5 |
| | Antwerp | 91.8 | 0.89 | 100 | 95.2 | 83.2 | 91.5 | 76.2 | 99.0 | 100 | 74.5 |
| | Leuven | 82.8 | 0.74 | 99.3 | 92.7 | 59.8 | 66.7 | 59.3 | 86.2 | 78.9 | 58.0 |
| | Hamburg | 75.9 | 0.67 | 99.1 | 93.7 | — | — | 53.8 | 70.5 | 93.3 | 28.0 |
| | Berlin | 77.8 | 0.68 | 96.5 | 96.8 | 57.9 | 51.5 | 49.8 | 80.5 | 90.7 | 44.0 |
| | Nantes | 72.3 | 0.69 | 80.7 | 77.2 | 42.9 | 41.0 | 71.5 | 84.0 | 77.0 | 77.0 |
| | Augsburg | 87.0 | 0.82 | 100 | 97.8 | — | — | 71.7 | 87.5 | 90.0 | 54.0 |
| | Paris | 71.2 | 0.69 | 96.4 | 72.2 | 68.3 | 88.5 | 45.1 | 71.0 | 70.1 | 51.8 |
| FCN | Wageningen | 75.2 | 0.61 | 94.7 | 96.1 | 32.3 | 69.9 | 62.5 | 42.5 | 62.1 | 47.0 |
| | Arnhem | 69.9 | 0.60 | 89.8 | 90.5 | — | — | 48.3 | 34.5 | 55.4 | 43.5 |
| | Brussels | 80.9 | 0.72 | 99.8 | 91.3 | 61.6 | 61.0 | 38.6 | 73.7 | 89.6 | 73.0 |
| | Antwerp | 88.3 | 0.85 | 100 | 93.0 | 76.6 | 85.0 | 69.9 | 97.5 | 97.2 | 68.5 |
| | Leuven | 79.0 | 0.68 | 98.8 | 93.0 | 55.0 | 63.5 | 51.7 | 75.7 | 69.9 | 47.5 |
| | Hamburg | 76.5 | 0.56 | 99.7 | 95.2 | — | — | 54.3 | 72.0 | 100 | 25.0 |
| | Berlin | 76.6 | 0.67 | 92.5 | 96.3 | 51.4 | 37.0 | 51.3 | 81.0 | 91.3 | 52.5 |
| | Nantes | 67.3 | 0.63 | 79.1 | 73.0 | 34.6 | 33.0 | 62.4 | 83.0 | 72.5 | 68.5 |
| | Augsburg | 84.6 | 0.77 | 99.2 | 98.2 | — | — | 67.4 | 81.5 | 81.0 | 47.0 |
| | Paris | 70.2 | 0.67 | 93.7 | 73.5 | 67.9 | 80.5 | 43.6 | 71.5 | 72.7 | 50.8 |
| RF | Wageningen | 68.3 | 0.52 | 83.4 | 92.0 | 29.8 | 65.8 | 50.0 | 37.0 | 96.6 | 24.3 |
| | Arnhem | 67.2 | 0.62 | 88.2 | 83.3 | 4.1 | 46.7 | 55.9 | 59.5 | 84.8 | 28.0 |
| | Brussels | 82.3 | 0.76 | 96.9 | 87.8 | 58.2 | 89.0 | 55.9 | 62.6 | 99.3 | 68.5 |
| | Antwerp | 78.3 | 0.76 | 92.9 | 78.7 | 51.4 | 80.0 | 70.6 | 87.5 | 100 | 66.5 |
| | Leuven | 75.3 | 0.65 | 97.8 | 87.0 | 37.8 | 67.7 | 54.1 | 76.8 | 74.6 | 42.5 |
| | Hamburg | 61.8 | 0.51 | 96.0 | 87.8 | — | — | 29.1 | 23.0 | 95.7 | 22.5 |
| | Berlin | 73.3 | 0.64 | 91.6 | 93.0 | 48.4 | 54.5 | 47.7 | 68.0 | 95.1 | 38.5 |
| | Nantes | 65.5 | 0.64 | 80.1 | 67.8 | 32.6 | 47.5 | 65.8 | 73.0 | 77.1 | 69.0 |
| | Augsburg | 81.1 | 0.78 | 98.7 | 90.8 | — | — | 70.1 | 84.5 | 94.2 | 48.5 |
| | Paris | 69.0 | 0.65 | 90.5 | 75.4 | 62.3 | 86.0 | 43.1 | 58.0 | 65.7 | 46.2 |

Table S6. Spatial transferability tests on the trained models for vertical height classification: Training samples from Danish cities used to train the models for mapping cities outside Denmark.

| Model | City | OA | Average F1 | Not built-up | | Low | | High | |
|---|---|---|---|---|---|---|---|---|---|
| | | | | UA | PA | UA | PA | UA | PA |
| DeepLab | Wageningen | 84.8 | 0.76 | 93.3 | 97.3 | 78.4 | 76.3 | 59.6 | 52.7 |
| | Arnhem | 69.3 | 0.61 | 80.3 | 96.5 | 64.2 | 43.0 | 44.6 | 41.3 |
| | Brussels | 84.4 | 0.77 | 99.8 | 95.3 | 89.9 | 56.3 | 46.1 | 100 |
| | Antwerp | 89.3 | 0.86 | 100 | 97.8 | 93.9 | 61.7 | 72.1 | 100 |
| | Leuven | 89.6 | 0.83 | 98.3 | 94.8 | 88.2 | 77.0 | 55.1 | 97.4 |
| | Hamburg | 83.6 | 0.79 | 97.3 | 94.8 | 63.9 | 86.0 | 83.4 | 58.7 |
| | Berlin | 85.8 | 0.82 | 95.5 | 98.7 | 68.1 | 84.0 | 88.1 | 61.7 |
| | Nantes | 71.9 | 0.69 | 82.9 | 83.2 | 45.7 | 63.7 | 96.1 | 57.7 |
| | Augsburg | 82.4 | 0.76 | 99.5 | 99.0 | 60.3 | 88.7 | 79.6 | 43.0 |
| | Paris | 73.2 | 0.72 | 94.0 | 77.4 | 52.3 | 74.3 | 75.4 | 65.3 |
| FCN | Wageningen | 84.1 | 0.76 | 97.7 | 95.9 | 78.6 | 73.3 | 48.9 | 61.6 |
| | Arnhem | 67.8 | 0.60 | 85.0 | 89.0 | 50.4 | 62.7 | 46.2 | 30.7 |
| | Brussels | 87.8 | 0.81 | 100 | 96.0 | 90.5 | 67.0 | 53.4 | 99.2 |
| | Antwerp | 88.9 | 0.87 | 100 | 90.7 | 79.9 | 74.3 | 79.6 | 100 |
| | Leuven | 91.7 | 0.87 | 99.3 | 93.8 | 86.2 | 87.7 | 66.7 | 90.9 |
| | Hamburg | 79.0 | 0.73 | 99.3 | 91.3 | 54.7 | 95.7 | 91.9 | 37.7 |
| | Berlin | 79.4 | 0.72 | 99.0 | 96.2 | 55.8 | 88.7 | 78.6 | 36.7 |
| | Nantes | 63.4 | 0.58 | 84.5 | 77.2 | 37.8 | 71.7 | 100 | 27.7 |
| | Augsburg | 80.6 | 0.72 | 100 | 97.5 | 56.5 | 96.7 | 90.2 | 30.7 |
| | Paris | 59.2 | 0.48 | 96.1 | 75.2 | 40.1 | 90.0 | 25.0 | 2.7 |
| RF | Wageningen | 72.2 | 0.51 | 82.4 | 93.2 | 58.5 | 64.3 | 11.1 | 1.8 |
| | Arnhem | 68.5 | 0.60 | 82.3 | 85.8 | 50.5 | 79.0 | 66.7 | 23.3 |
| | Brussels | 81.8 | 0.75 | 97.0 | 91.0 | 74.4 | 59.0 | 50.9 | 93.3 |
| | Antwerp | 81.4 | 0.79 | 97.0 | 81.3 | 62.7 | 63.3 | 75.9 | 99.7 |
| | Leuven | 84.4 | 0.76 | 97.8 | 87.8 | 72.0 | 82.3 | 53.7 | 66.2 |
| | Hamburg | 75.7 | 0.70 | 94.3 | 88.3 | 52.1 | 87.0 | 85.4 | 39.0 |
| | Berlin | 78.7 | 0.72 | 96.3 | 94.8 | 55.1 | 87.0 | 84.4 | 38.0 |
| | Nantes | 65.1 | 0.64 | 84.1 | 69.5 | 39.8 | 71.0 | 89.3 | 50.3 |
| | Augsburg | 76.6 | 0.66 | 97.2 | 93.3 | 52.9 | 98.3 | 97.0 | 21.3 |
| | Paris | 67.9 | 0.65 | 89.6 | 76.8 | 45.4 | 81.0 | 85.5 | 39.3 |

 

\end{document}